\documentclass[prd,aps,notitlepage,twocolumn,longbibliography,nofootinbib]{revtex4-2} 
\usepackage{amsfonts,amssymb,amsmath,mathrsfs,graphicx,xcolor,bm,mathtools,physics,orcidlink,url,revsymb}
\definecolor{ultramarine}{rgb}{0.07, 0.04, 0.56}
\definecolor{cadmiumgreen}{rgb}{0.0, 0.42, 0.24}
\definecolor{indigo(dye)}{rgb}{0.0, 0.25, 0.42}
\usepackage{hyperref}
\hypersetup{
colorlinks=true,
citecolor=ultramarine,
linkcolor=cadmiumgreen,
urlcolor=indigo(dye),
}

\newcommand{\f}[2]{\frac{#1}{#2}}
 
\newcommand{\mk}[1]{\left( #1 \right)}

\newcommand{\be}{\begin{equation}}  
\newcommand{\ee}{\end{equation}}

\newcommand{\ve}{\varepsilon}
\renewcommand{\Re}{{\rm Re}\,}
\renewcommand{\Im}{{\rm Im}\,}

\newcommand{\HeunC}{\mathrm{HeunC}}
\newcommand{\res}{\operatorname*{Res}}

\begin{document}

\title{
Pole structure of the Kerr Green's function
}

\author{Hayato Motohashi\,\orcidlink{0000-0002-4330-7024}}
\affiliation{Department of Physics, Tokyo Metropolitan University, 1-1 Minami-Osawa, Hachioji, Tokyo 192-0397, Japan}

\author{Yuto Suichi\,\orcidlink{0009-0001-0591-0869}}
\affiliation{Department of Physics, Tokyo Metropolitan University, 1-1 Minami-Osawa, Hachioji, Tokyo 192-0397, Japan}


\begin{abstract}
We investigate the pole structure of Kerr black-hole perturbations in the frequency domain, focusing on the building blocks of the Green's function for the radial Teukolsky equation: the homogeneous radial solutions, the connection coefficients, and the Green's function itself.
We show that the homogeneous solutions and the local connection coefficients develop simple poles at the Matsubara frequencies, thereby establishing the Matsubara pole structure explicitly within the Teukolsky formalism for asymptotically flat Kerr black holes.
At the level of the local fixed-sector connection formula, the explicit Matsubara-pole factors cancel in the ratio of connection coefficients entering a decomposed Green's function contribution.
We also identify higher-order zero-frequency singularities in the decomposed Green's function contributions, which scale as $\omega^{-2l-1}$ and cancel collectively in the total radial Green's function.
These results clarify how Matsubara poles and sectoral zero-frequency singularities arise in the Teukolsky formalism and provide a frequency-domain foundation for understanding prompt response in time-domain ringdown waveforms in Kerr spacetime.
\end{abstract}

\maketitle  

\section{Introduction}
\label{sec:intro}

Black-hole ringdown has become a central target of gravitational-wave black-hole spectroscopy~\cite{Kokkotas:1999bd,Berti:2009kk,Berti:2025hly}.
At the level of linear perturbation theory, it is conventionally understood in terms of poles and branch cuts of the frequency-domain Green's function.
In the time domain, these structures underlie the familiar quasinormal-mode (QNM) ringing and Price-law tails, and therefore provide the basic analytic framework for linear black-hole response~\cite{Price:1971fb,Leaver:1986gd,Andersson:1996cm}.
By contrast, the early-time, or prompt, response has only recently begun to be understood in comparable detail.

The prompt-response regime is important for several reasons.
First, the signal is typically stronger at earlier times than in the later ringdown, so improving its theoretical description is directly relevant for gravitational-wave phenomenology.
Second, it characterizes the transition between the source-dependent merger signal and the QNM-dominated regime, thereby affecting the choice of ringdown start time and the interpretation of QNM excitation amplitudes.
Third, a precise understanding of the prompt part of the linear response is indispensable for identifying genuinely nonlinear features in the near-merger waveform: without first determining what is already encoded in the linear Green's function, one cannot unambiguously isolate nonlinear effects.

Recent progress indicates that prompt response can be associated with several distinct kinds of singular structures, depending on the asymptotic geometry and on how the Green's function is decomposed.
For short-range effective potentials, such as the P\"oschl--Teller model and the Schwarzschild--de Sitter case, the prompt response can be described by Matsubara, or Euclidean, modes on the imaginary axis, and the corresponding mode sum converges in the early-time region~\cite{Kuntz:2025gdq,Arnaudo:2025uos}.
This terminology is borrowed from finite-temperature field theory, where Matsubara frequencies are fixed by the inverse temperature~\cite{LeBellac1996}. 
In the black-hole context, the relevant temperature is the Hawking temperature.
Here, the terminology refers to the characteristic frequency pattern set by the Hawking temperature, rather than to an explicit construction of a thermal Euclidean Green's function.
In the asymptotically flat limit, another family of imaginary-axis poles present in the Schwarzschild--de Sitter problem, distinct from the Matsubara modes organizing the prompt response, condenses into the branch-cut structure responsible for Price-law behavior~\cite{Arnaudo:2025kit}.

For asymptotically flat Schwarzschild spacetime, a complementary picture has emerged.
The branch-cut structure of this problem has a long history~\cite{Leaver:1986gd,Leung:2003ix,Casals:2013mpa,Casals:2012ng,Casals:2015nja}, and recent work has substantially refined the Green's function perspective.
On the time-domain side, recent studies have developed direct and regularized calculations of Schwarzschild Green's functions, clarifying how singular features of wave propagation appear in the time-domain response~\cite{Aruquipa:2026tga,Aruquipa:2026kqk}.
On the frequency-domain side, several analyses have shown that the prompt response and the late-time tail can be organized in terms of zero-frequency poles, imaginary-axis branch cuts, and causal decompositions of the Green's function~\cite{DeAmicis:2026tus,Su:2026fvj,Rosato:2026moe}.
In particular, the prompt response has been related to the zero-frequency structure rather than to a large-frequency arc, while branch-cut contributions naturally account for prompt and tail components depending on the contour and causal regime.
These developments strongly suggest that prompt response is controlled by nontrivial frequency-domain singular structures and is not a negligible correction to ringdown.

Related time-domain studies have also discussed source-dependent early-time features under names such as horizon modes~\cite{Mino:2008at,Zimmerman:2011sv}, redshift mode~\cite{DeAmicis:2025xuh}, and direct waves~\cite{Oshita:2025tbc}.
In addition, analyses of plunging sources have shown that prompt components can be quantitatively important in waveform reconstruction~\cite{Ma:2026qbq}.
These phenomena may be physically relevant to the early-time signal and, in some cases, provide an efficient description of waveform components beyond the standard QNM-plus-tail picture.
At the same time, such source-dependent signals are obtained by convolving the Green's function with the source or initial data.
Therefore, clarifying the source-independent singular structure of the Green's function itself is a necessary step toward understanding how these early-time components arise in concrete waveforms.

From the viewpoint of the Green's function and its role in ringdown, the current picture is that Matsubara poles, branch cuts, and zero-frequency singularities may all be relevant to prompt response, with the most useful description depending on the asymptotic structure of the spacetime, the contour decomposition, and the physical regime under consideration.
What remains largely unexplored, in particular, is the corresponding fixed-sector analytic structure for rotating black holes.

The Kerr case is qualitatively richer.
In rotating backgrounds, the frequencies naturally associated with thermal, or Matsubara-like, structures are shifted away from the imaginary axis by the horizon angular velocity, as is familiar from Kerr/conformal field theory (CFT) and hidden-conformal-symmetry analyses~\cite{Chen:2010ni}.
However, to the best of our knowledge, in asymptotically flat subextremal Kerr there is still no first-principles derivation of how such Matsubara frequencies appear, or fail to appear, in the building blocks of the Green's function constructed from the Teukolsky equation.
Likewise, the zero-frequency singularities in Kerr have not been derived directly from the Teukolsky equation in a form suitable for analyzing their role in the Green's function.

In this paper, we analyze the analytic structure of the basic building blocks of the Kerr Green's function by working directly with the Teukolsky equation: the homogeneous radial solutions, the connection coefficients, and the Green's function assembled from them.
In particular, we clarify the Matsubara pole structure and the zero-frequency singularities of the homogeneous radial Teukolsky solutions, the associated connection coefficients, and the decomposed Green's function contributions.
We use the confluent Heun formulation to expose the local analytic structure of the homogeneous solutions and connection problem, and the Mano--Suzuki--Takasugi (MST) method~\cite{Mano:1996vt,Mano:1996gn,Sasaki:2003xr} to control the low-frequency behavior.
Our focus is on a first-principles characterization of the fixed-sector frequency-domain pole structure, which provides the analytic foundation for future studies of prompt response in rotating black holes.

The rest of the paper is organized as follows.
In \S\ref{sec:teu}, we review the radial Teukolsky equation and fix our notation.
In \S\ref{sec:HeunC}, we recast it in the form of the confluent Heun equation; basic properties of the confluent Heun functions are summarized in Appendix~\ref{app:HeunCbasics}, and the pole/residue formulae used in our analysis are derived in Appendix~\ref{app:poleres}.
In \S\ref{sec:pole}, we clarify the Matsubara pole structure and zero-frequency behavior of the homogeneous solutions, the connection coefficients, and the Green's function for Kerr black-hole perturbations within the Teukolsky formalism, using the confluent Heun formulation together with the MST results summarized in Appendix~\ref{app:mst}.
For comparison, Appendix~\ref{app:RWMST} analyzes the corresponding structures in the Regge--Wheeler (RW) formalism for Schwarzschild black holes using the MST representation.
\S\ref{sec:con} is devoted to conclusions and discussion.
We adopt geometric units $G=c=1$ throughout.

\section{Teukolsky equation}
\label{sec:teu}

In this section, we briefly review the Teukolsky formalism for Kerr black-hole perturbations, mostly following Ref.~\cite{Sasaki:2003xr}.
The radial Teukolsky equation is given by
\begin{align}\label{eq:radTeu}
&\Biggl[\Delta^{-s}\frac{\dd }{\dd r}\biggl(\Delta^{s+1}\frac{\dd}{\dd r}\biggr) + \frac{K^2-2is(r-M)K}{\Delta} \notag\\
&+4is\omega r-\lambdabar\Biggr]{}_sR_{lm} = {}_sT_{lm},
\end{align}
where 
\begin{align}
K&=(r^2+a^2)\omega-ma, \\
\Delta&=r^2-2Mr+a^2,\notag\\
&=(r-r_+)(r-r_-), \\
r_\pm&=M\pm\sqrt{M^2-a^2}, \\
\lambdabar &= {}_sA_{lm} - 2m a\omega + a^2\omega^2,
\end{align}
where ${}_sA_{lm}$ is the separation constant, and ${}_sT_{lm}$ is the source term.

The radial Teukolsky equation can be solved by the Green's function method, where the Green's function is constructed from two linearly independent homogeneous solutions.
Although any pair of linearly independent homogeneous solutions may be used, a conventional choice is the ``in'' solution ${}_sR^\mathrm{in}_{lm}$, which satisfies
\be {}_sR^\mathrm{in}_{lm}(r) \to
\begin{cases}
\Delta^{-s}e^{-ip r_*}, & (r\to r_+), \\ 
{}_sB^\mathrm{ref}_{lm}r^{-1-2s}e^{i\omega r_*} + {}_sB^\mathrm{inc}_{lm}r^{-1}e^{-i\omega r_*}, & (r\to \infty),
\end{cases}
\label{eq:Rinasymp}
\ee
and the ``up'' solution ${}_sR^\mathrm{up}_{lm}$, which satisfies 
\be {}_sR^\mathrm{up}_{lm}(r) \to
\begin{cases}
{}_sC^\mathrm{up}_{lm}e^{ip r_*} + {}_sC^\mathrm{ref}_{lm}\Delta^{-s}e^{-ip r_*}, & (r\to r_+), \\ 
r^{-1-2s}e^{i\omega r_*}, & (r\to \infty),
\end{cases}
\label{eq:Rupasymp}
\ee
where 
\be p = \omega -m\Omega_+, \quad \Omega_\pm =\frac{a}{2Mr_\pm}, \ee
and ${}_sB^\mathrm{inc/ref}_{lm}$ and ${}_sC^\mathrm{up/ref}_{lm}$ are complex asymptotic amplitudes.
Note that here we normalize ${}_sR^\mathrm{in}_{lm}$ and ${}_sR^\mathrm{up}_{lm}$ so that the transmission amplitudes are unity; these transmission amplitudes are denoted by ${}_s\bar B^\mathrm{trans}_{lm}$ and ${}_s\bar C^\mathrm{trans}_{lm}$ in the literature~\cite{Sasaki:2003xr}.
The tortoise coordinate $r_*$ is defined by $\dd r_*/\dd r=(r^2+a^2)/\Delta$, and we fix the integration constant by the conventional choice
\be \label{eq:tortoise} r_* = r + \frac{2Mr_+}{r_+ - r_-}\ln\frac{r-r_+}{2M} - \frac{2Mr_-}{r_+ - r_-}\ln\frac{r-r_-}{2M}. \ee

The other two homogeneous solutions are defined analogously:
\begin{align}
{}_sR^\mathrm{out}_{lm}&\to e^{ip r_*}, & (r\to r_+), \\
{}_sR^\mathrm{down}_{lm}&\to r^{-1}e^{-i\omega r_*}, & (r\to \infty),
\end{align}
and can be constructed from the corresponding spin-flipped solutions ${}_{-s}R^\mathrm{in}_{lm}$ and ${}_{-s}R^\mathrm{up}_{lm}$ through the conjugation relations
\begin{align}
\label{eq:Routin} {}_sR^\mathrm{out}_{lm}(\omega,r)&= \Delta^{-s}(r) \bigl[{}_{-s}R^\mathrm{in}_{lm}(\omega^*,r)\bigr]^*,\\
\label{eq:Rdownup} {}_sR^\mathrm{down}_{lm}(\omega,r)&= \Delta^{-s}(r) \bigl[{}_{-s}R^\mathrm{up}_{lm}(\omega^*,r)\bigr]^*,
\end{align}
where we have written the $\omega$ dependence explicitly.

From now on, for simplicity, we omit the subscripts $s$ and $lm$ whenever no confusion arises.
At infinity, $R^\mathrm{up}$ and $R^\mathrm{down}$ form the outgoing and ingoing basis, respectively. 
Therefore, the ``in'' solution can be decomposed as
\begin{align}
\label{eq:Rindecomp}
R^\mathrm{in}(r)=B^\mathrm{ref}R^\mathrm{up}(r)+B^\mathrm{inc}R^\mathrm{down}(r).
\end{align}
In this sense, the asymptotic amplitudes $B^\mathrm{inc}$ and $B^\mathrm{ref}$ may be regarded as connection coefficients between the horizon-adapted solution and the asymptotic basis at infinity.

As mentioned above, the Green's function is constructed from $R^\mathrm{in}$ and $R^\mathrm{up}$ as
\begin{align}\label{eq:Green}
G(r,r') 
&= \frac{\Delta^s(r')}{2i\omega B^\mathrm{inc}} 
\begin{cases}
R^\mathrm{up}(r)R^\mathrm{in}(r'), & (r>r'), \\
R^\mathrm{in}(r)R^\mathrm{up}(r'), & (r<r').
\end{cases}
\end{align}
The inhomogeneous solution of the radial Teukolsky equation~\eqref{eq:radTeu} in the frequency domain is then given by
\begin{align}
R(r) &= \int_{r_+}^{\infty}\dd r' G(r,r') T(r') .
\end{align}

Let us focus on the case $r>r'$.
Substituting \eqref{eq:Rindecomp} into \eqref{eq:Green}, we decompose the Green's function as
\begin{align}
G(r,r') = G^{(+)}(r,r') + G^{(-)}(r,r') ,
\end{align}
where
\begin{equation}\begin{split} \label{eq:Gpm}
G^{(+)}(r,r') &=
\frac{B^\mathrm{ref}}{2i\omega B^\mathrm{inc}} 
\Delta^s(r') R^\mathrm{up}(r') R^\mathrm{up}(r) , \\
G^{(-)}(r,r') &=
\frac{1}{2i\omega } 
\Delta^s(r') R^\mathrm{down}(r') R^\mathrm{up}(r) .
\end{split}\end{equation}
This decomposition is useful for analyzing separate contributions to the Green's function and for discussing their causal interpretation~\cite{Andersson:1996cm,Kuntz:2025gdq,Arnaudo:2025uos,Su:2026fvj}.
Distinguishing these two contributions is crucial for clarifying the time-domain signal in the early-time regime,
$|r_*-r'_*|<t-t'<|r_*+r'_*|$, where they enter the contour analysis in different ways.

\section{Confluent Heun equation}
\label{sec:HeunC}

Both the angular and radial Teukolsky equations can be transformed into confluent Heun equations.
Here we focus on the radial equation.
We introduce the new independent variable
\be \label{eq:zr} z = \f{r - r_-}{r_+ - r_-}. \ee
We then redefine the radial function as
\be \label{eq:Rpsi} R(r) = z^{\nu_1} (z - 1)^{\nu_2} e^{\nu_3 z} \psi(z), \ee
where 
\begin{equation}\begin{split}
\nu_1 &= \f{i(\omega - m  \Omega_-)}{2 \kappa_-} , \quad
\nu_2 = -s - \f{i(\omega - m  \Omega_+)}{2 \kappa_+} , \\
\nu_3 &= \f{2 i \kappa_+ a \omega}{\Omega_+} .
\end{split}\end{equation}
The surface gravities are
\be \kappa_\pm=\f{1}{4M}-M\Omega_\pm^2 
=\frac{r_\pm-r_\mp}{4Mr_\pm}. \ee
Thus, $\kappa_+>0$, whereas $\kappa_-<0$ in this convention.
The Hawking temperature is determined by the outer-horizon surface gravity,
\be T_\mathrm{H}=\frac{\kappa_+}{2\pi}. \ee

This transformation maps the regular singular points $r=r_-,r_+$ and the irregular singular point $r=\infty$ to $z=0,1$, and $\infty$, respectively.
Then $\psi(z)$ satisfies a confluent Heun equation,
\begin{align} \label{eq:HeunCeq}
\psi'' + \mk{ \f{\gamma}{z} + \f{\delta}{z-1} + \epsilon } \psi' + \f{\alpha z - q}{z(z-1)} \psi = 0 , 
\end{align}
where 
\begin{equation}\begin{split}\label{eq:Heunparams}
q &= {}_sA_{lm} - (\nu_1 + \nu_2)(1 + 2 s + \nu_1 + \nu_2) \\
  &~~~+ (1 + 2  s + 2 \nu_1) \nu_3 
    + \frac{(r_+^2 + 3 r_- r_+ + 3 r_-^2)\nu_3^2}{(r_+ - r_-)^2}, \\
\alpha &= 2 \nu_3 (1 + 3  s + 2  \nu_2),\quad
\gamma = 1 + s + 2\nu_1 , \\
\delta &= 1 + s + 2\nu_2, \quad
\epsilon = 2\nu_3.
\end{split}\end{equation}

Around each singular point $z=z_s$, we can construct two linearly independent solutions $\psi_j^{(z_s)}$ with $j=1,2$ in terms of confluent Heun functions.
The four solutions are written as
\begin{equation}\begin{split} \label{eq:InOutUpDown}
\psi_\mathrm{in}(z) &= N_\mathrm{in} \psi_1^{(1)} , \quad 
\psi_\mathrm{out}(z) = N_\mathrm{out} \psi_2^{(1)}, \\
\psi_\mathrm{up}(z) &= N_\mathrm{up} \psi_1^{(\infty)}, \quad 
\psi_\mathrm{down}(z) = N_\mathrm{down} \psi_2^{(\infty)} ,
\end{split}\end{equation}
where the normalization constants are
\begin{equation}\begin{split}
N_\mathrm{in} &=(r_+-r_-)^{-2s}\Big(\frac{2M}{r_+-r_-}\Big)^{2iMp}e^{-ipr_+}e^{-i\omega(r_+-r_-)}, \\ 
N_\mathrm{out} &=\Big(\frac{2M}{r_+-r_-}\Big)^{-2iMp}e^{ipr_+}e^{-i\omega(r_+-r_-)}e^{-i\pi(1-\delta)} , \\
N_\mathrm{up} &=(r_+-r_-)^{-1-2s}\Big(\frac{2M}{r_+-r_-}\Big)^{ -2iM\omega}e^{i\omega r_-} , \\ 
N_\mathrm{down} &=(r_+-r_-)^{-1}\Big(\frac{2M}{r_+-r_-}\Big)^{2iM\omega}e^{-i \omega r_-} .
\end{split}\end{equation}
Here, $\psi_j^{(1)}$ and $\psi_j^{(\infty)}$ are provided in Appendix~\ref{app:HeunCbasics} and are written in terms of $\HeunC(q,\alpha,\gamma,\delta,\epsilon;z)$ and $\HeunC_\infty (q,\alpha,\gamma,\delta,\epsilon; z)$, respectively.
Following the notation in Ref.~\cite{Bonelli:2022ten}, $\HeunC(q,\alpha,\gamma,\delta,\epsilon;z)$ denotes the standard confluent Heun function whose expansion around $z=0$ is given by
\begin{align} 
\HeunC(q,\alpha,\gamma,\delta,\epsilon;z) = 1-\f{q}{\gamma}z + \order{z^2} ,
\end{align}
and $\HeunC_\infty (q,\alpha,\gamma,\delta,\epsilon; z)$ denotes a different function whose asymptotic expansion around the irregular singularity $z=\infty$ is given by
\begin{align} 
&\HeunC_\infty(q,\alpha,\gamma,\delta,\epsilon;z^{-1}) \\
&= 1+\f{\alpha^2-(\gamma+\delta-1)\alpha\epsilon+(\alpha-q)\epsilon^2}{\epsilon^3}z^{-1} + \order{z^{-2}} . \notag
\end{align}
Explicit series expansions of these functions and their key properties are summarized in Appendix~\ref{app:HeunCbasics}.
We also derive pole and residue formulae for the confluent Heun function in Appendix~\ref{app:poleres}.

Alternatively, the homogeneous solutions may be represented in the MST formalism as infinite series of confluent hypergeometric functions~\cite{Mano:1996vt,Mano:1996gn,Sasaki:2003xr}.
This representation is often advantageous because it yields convergent series with a wider domain of validity than a local asymptotic expansion and is well suited to analytic continuation and low-frequency expansions (see Appendix~\ref{app:mst}).

\section{Pole and zero structure}
\label{sec:pole}

\begin{figure*}[t]
\centering
\includegraphics[width=1\linewidth]{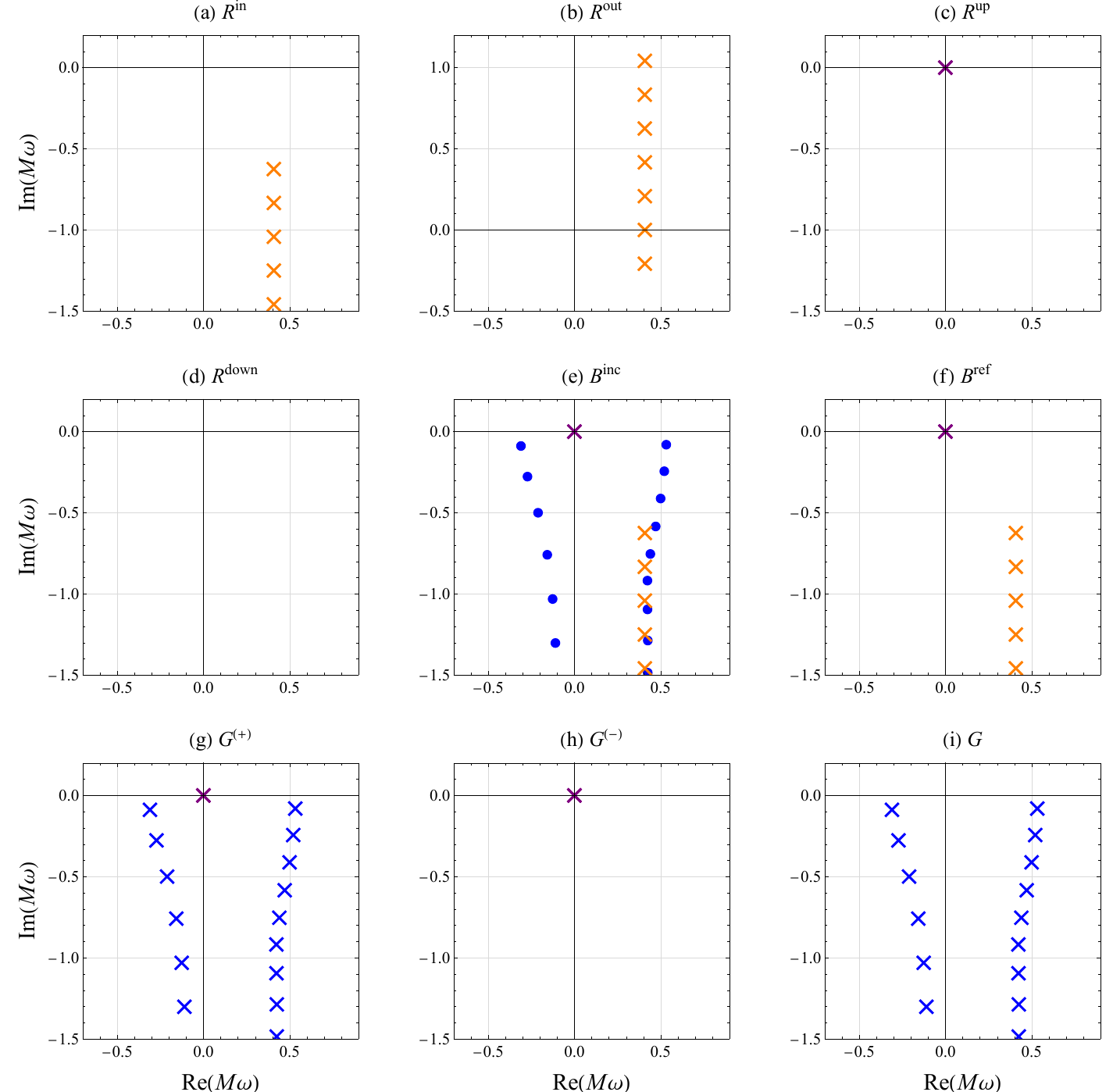}
\caption{Pole (crosses) and zero (dots) structure of the Green's function and its building blocks in the complex frequency plane for $(s,l,m)=(-2,2,2)$ and $a/M=0.7$.
The QNM (blue), Matsubara (orange), and zero-frequency (purple) structures are shown, omitting the branch-cut structure and any additional poles or zeros that may be present. 
The QNM frequencies are taken from the public dataset~\cite{motohashi_2024_12696857} associated with Ref.~\cite{Motohashi:2024fwt}.
The top- and middle-row panels show the pole and zero structures of the homogeneous solutions ${}_sR^{\mathrm{in}}_{lm}$, ${}_sR^{\mathrm{out}}_{lm}$, ${}_sR^{\mathrm{up}}_{lm}$, ${}_sR^{\mathrm{down}}_{lm}$ and the connection coefficients ${}_sB^{\mathrm{inc}}_{lm}$, ${}_sB^{\mathrm{ref}}_{lm}$.
The bottom-row panels display the pole structures of the contributions ${}_sG^{(+)}_{lm}$ and ${}_sG^{(-)}_{lm}$, and the Green's function ${}_sG_{lm}$ itself.
}
\label{fig:Graph_ComplexPlane}
\end{figure*}

Using the formulation developed in the previous sections, we now analyze the analytic structure of the basic building blocks of the Kerr Green's function: the homogeneous solutions (\S\ref{ssec:homo}), the connection coefficients (\S\ref{ssec:conn}), and the Green's function itself (\S\ref{ssec:Green}).

As a guide to the discussion, we first present in Fig.~\ref{fig:Graph_ComplexPlane} the pole and zero structure obtained for a representative parameter set with $(s,l,m)=(-2,2,2)$ and $a/M=0.7$.
In Fig.~\ref{fig:Graph_ComplexPlane}, we display only the QNM (blue), Matsubara (orange), and zero-frequency (purple) structures relevant to the present discussion.
For clarity, the branch-cut structure is omitted, as are any additional poles or zeros that may be present.

While the building blocks possess explicit Matsubara poles, these poles are absent in the plotted fixed-sector quantities $G^{(\pm)}$.
In this sense, the ratio entering $G^{(+)}$ exhibits an explicit factor cancellation: the Matsubara-pole factors present in both $B^{\mathrm{ref}}$ and $B^{\mathrm{inc}}$ cancel in the quotient.
The zero-frequency structure is more subtle: the decomposed contributions $G^{(+)}$ and $G^{(-)}$ possess higher-order zero-frequency singularities, whereas these singularities cancel in the total Green's function $G$.

Below, we explain how these structures arise analytically and verify numerically that the corresponding singularities are indeed realized.
For the numerical checks, we make use of the Black Hole Perturbation Toolkit~\cite{BHPToolkit}.

Some of these structures persist in other master-variable formulations, whereas others are tied to the Teukolsky variable and its normalization.
In particular, the RW comparison in Appendix~\ref{app:RWMST} shows that, although the spin-weight-dependent form of the local Matsubara frequencies and of the homogeneous-solution pole orders is reorganized in the RW variable, the explicit Matsubara-pole factors again cancel in the decomposed Green's function contribution, and the Green's function-level low-frequency order counting agrees with the Teukolsky result.

\subsection{Homogeneous solutions}
\label{ssec:homo}

We begin with the homogeneous solutions, $R^\mathrm{in}$, $R^\mathrm{out}$, $R^\mathrm{up}$, and $R^\mathrm{down}$, which form the first building block of the Kerr Green's function. 
From the poles of the confluent Heun function summarized in Appendix~\ref{apps:MMRinRout}, $R^\mathrm{in}$ develops first-order singularities at $\delta=-k$ with $k=0,1,2,\ldots$.
This condition defines the frequencies
\be \label{eq:MMfreq} {}_s\omega_{mk}^{(-)} = m \Omega_+ - i (1 - s + k) \kappa_+, \quad (k=0,1,2,\ldots). \ee
The same pole condition can also be derived independently in the MST formalism (see Appendix~\ref{apps:MMRinRout-mst}).

The frequencies~\eqref{eq:MMfreq} are precisely the Matsubara frequencies of Kerr black holes.
This reproduces, directly from the radial Teukolsky equation, the frequency pattern previously suggested in the extremal Kerr/CFT context~\cite{Chen:2010ni}, and extends it to a first-principles analytic derivation for generic subextremal Kerr and arbitrary spin weight $s$, without using a near-extremal approximation.

Their spacing in the imaginary direction is set by $\kappa_+=2\pi T_\mathrm{H}$, while their real part is shifted by $m\Omega_+$ due to the rotation of the horizon.
Equivalently, the natural thermal frequency near the horizon is measured with respect to the co-rotating Killing vector $\partial_t+\Omega_+\partial_\phi$, so a mode with azimuthal number $m$ appears through the combination $\omega-m\Omega_+$.
This shift is analogous to the frequency shift produced by a chemical potential in finite-temperature field theory~\cite{LeBellac1996}.
It is worth stressing, however, that our derivation does not rely on constructing a thermal correlator; rather, the radial Teukolsky equation itself produces a pole pattern with the finite-temperature form dictated by $T_\mathrm{H}$ and $\Omega_+$.

Similarly, from \eqref{eq:InOutUpDown} we note that $R^\mathrm{out}$ develops first-order singularities at $2-\delta=-k$ with $k=0,1,2,\ldots$, which defines
\be \label{eq:MMfreqPlus} {}_s\omega_{mk}^{(+)} = m \Omega_+ + i (1 + s + k) \kappa_+, \quad (k=0,1,2,\ldots). \ee
These frequencies form the counterpart of the Matsubara branch associated with the ``out'' solution.
They are also consistent with \eqref{eq:Routin}, i.e., 
\be {}_s\omega_{mk}^{(+)} = {}_{-s}\omega_{mk}^{(-)*}. \ee

Unlike the QNM frequencies ${}_s\omega_{lmn}$, the Matsubara frequencies ${}_s\omega_{mk}^{(\pm)}$ identified here are independent of the angular quantum number $l$.
At the same time, their detailed form should be understood within the Teukolsky-variable representation and its normalization: in this formulation they depend linearly on the spin weight $s$, and the superscripts $(\pm)$ are merely labels adapted to the local horizon solutions $R^\mathrm{in}$ and $R^\mathrm{out}$.
They should not be interpreted as a universal division into lower- and upper-half-plane Matsubara sectors; indeed, depending on $s$, the frequencies in the set~\eqref{eq:MMfreqPlus} need not all have the same sign of the imaginary part, as exemplified in Fig.~\ref{fig:Graph_ComplexPlane}(b) for $s=-2$.
These features should be distinguished from formulations based on RW master variables [see \eqref{eq:RW-MM-candidate} and \eqref{eq:RW-MM-candidate-plus}] or asymptotically de Sitter problems in the literature, where the corresponding Matsubara structure may be organized differently.

Using the confluent Heun analysis, the residues of $R^\mathrm{in}$ and $R^\mathrm{out}$ can be analytically obtained as
\begin{align}
\label{eq:Rin-residue}
&\res_{\omega={}_s\omega^{(-)}_{mk}}{}_sR_{lm}^{\mathrm{in}}\notag\\
&=
\left.
i\kappa_+ N_{\mathrm{in}}z^{\nu_1}(z-1)^{\nu_2}e^{\nu_3 z}
\res_{\delta=-k}\psi_1^{(1)}
\right|_{\omega={}_s\omega^{(-)}_{mk}},
\\[5pt]
\label{eq:Rout-residue}
&\res_{\omega={}_s\omega^{(+)}_{mk}}{}_sR_{lm}^{\mathrm{out}}\notag\\
&=
\left.
-i\kappa_+ N_{\mathrm{out}}z^{\nu_1}(z-1)^{\nu_2}e^{\nu_3 z}
\res_{2-\delta=-k}\psi_2^{(1)}
\right|_{\omega={}_s\omega^{(+)}_{mk}}.
\end{align}
Here, the residues of $\psi_1^{(1)}$ and $\psi_2^{(1)}$ are analytically written down in terms of the confluent Heun functions as given in \eqref{eq:psi1-residue} and \eqref{eq:psi2-residue}, respectively.
The same residues can be obtained semianalytically in the MST formalism as
\begin{align} \label{eq:Rin-residue-MST}
\res_{\omega={}_s\omega^{(-)}_{mk}}{}_sR^\mathrm{in}_{lm}
&= \left.
i\kappa_+\frac{(-1)^k}{k!}
\frac{{}_sR^\mathrm{in}_{lm}}{\Gamma(\delta)}
\right|_{\omega={}_s\omega^{(-)}_{mk}} , \\
\label{eq:Rout-residue-MST}
\res_{\omega={}_s\omega^{(+)}_{mk}}{}_sR^\mathrm{out}_{lm}
&= \left.
-i\kappa_+\frac{(-1)^k}{k!}
\frac{{}_sR^\mathrm{out}_{lm}}{\Gamma(2-\delta)}
\right|_{\omega={}_s\omega^{(+)}_{mk}} .
\end{align}

By contrast, the transmission-normalized solutions $R^\mathrm{in}_{lm}$ and $R^\mathrm{out}_{lm}$ are regular at $\omega=0$ for fixed $r>r_+$, as follows from the MST analysis in Appendix~\ref{apps:MMRinRout-mst}.

On the other hand, $R^\mathrm{up}$ exhibits a distinct singular structure at zero frequency.
For nonzero spin weight $s\ne 0$, as demonstrated in Appendix~\ref{app:mst}, the low-frequency expansion of the MST representation gives, for fixed $r>r_+$, 
\begin{align}
\label{eq:Ruppole}
{}_sR^\mathrm{up}_{lm}(r)
&= {}_sQ_{lm}(r) \omega^{s-l} + \order{\omega^{s-l+1}},
\end{align}
where the leading-order coefficient ${}_sQ_{lm}(r)$ is explicitly given in \eqref{eq:Rup-leading-coeff-2}.
This singularity is different in origin from the Matsubara poles discussed above: it arises from the degenerate low-frequency limit of the MST expansion rather than from the pole condition $\delta=-k$ of the local Heun solution.

In particular, for the gravitational lowest multipole $(s,l)=(-2,2)$, $R^\mathrm{up}$ has a fourth-order pole as
\begin{align}\label{eq:Rup-leading-s22-series}
{}_{-2}R^\mathrm{up}_{2m}(r) &= {}_{-2}Q_{2m}(r)\omega^{-4}+\order{\omega^{-3}},
\end{align}
with
\begin{align}
{}_{-2}Q_{2m}(r)
&=
\frac{3}{2}\,
\left(\frac{r-r_+}{r-r_-}\right)^{\frac{i m \chi}{2\kappa}}
\frac{(r-r_+)^2}{(r-r_-)^3} \notag\\
&\quad\times {}_2F_1\left(5,3+\frac{i m\chi}{\kappa};6;\frac{r_+-r_-}{r-r_-}\right),
\end{align}
where $\chi=a/M$ and $\kappa=\sqrt{1-\chi^2}$.

The remaining solution $R^\mathrm{down}$ follows from the spin-flip conjugation relation \eqref{eq:Rdownup}.
Combining \eqref{eq:Rdownup} with \eqref{eq:Ruppole}, we obtain, for fixed $r>r_+$ [see also \eqref{eq:Rdown-leading-coeff}],
\begin{align}
\label{eq:Rdownpole}
{}_sR^\mathrm{down}_{lm}(r)
&= \Delta^{-s}(r) \bigl[{}_{-s}Q^*_{lm}(r) \omega^{-s-l} + \order{\omega^{-s-l+1}}\bigr].
\end{align}
For the gravitational lowest multipole $(s,l)=(-2,2)$, this gives $R^\mathrm{down}_{lm}= \order{\omega^0}$.
Thus, in this case, $R^\mathrm{down}_{lm}$ has neither Matsubara poles of the type found in $R^\mathrm{in}_{lm}$ and $R^\mathrm{out}_{lm}$ nor an analogous zero-frequency pole of the type found in $R^\mathrm{up}_{lm}$, as depicted in Fig.~\ref{fig:Graph_ComplexPlane}.

\begin{figure}[t]
\centering
\includegraphics[width=0.95\linewidth]{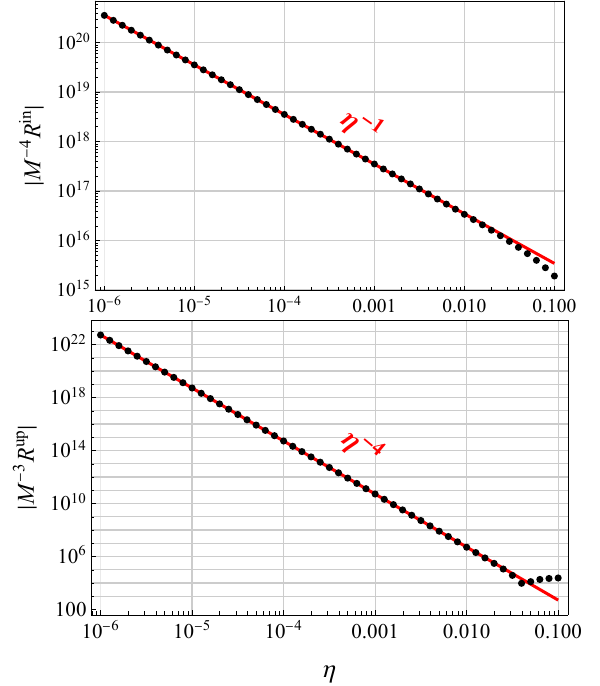}
\caption{
Magnitudes of ${}_sR^\mathrm{in}_{lm}(r)$ (top) and ${}_sR^\mathrm{up}_{lm}(r)$ (bottom) in the vicinity of their respective poles, shown as functions of $\eta=|M(\omega-\omega_\mathrm{pole})|$.
For the top panel, the pole is the fundamental Matsubara pole $\omega={}_s\omega^{(-)}_{m0}$; for the bottom panel, the pole is $\omega=0$.
Here, $(s,l,m)=(-2,2,2)$, $r=30M$, and $a/M=0.7$.
Black dots represent numerical results, while the red lines indicate reference scalings proportional to $\eta^{-1}$ and $\eta^{-4}$, respectively.
}
\label{fig:RinRup_MMand0}
\end{figure}

We now numerically verify the analytic results presented above.
We mainly focus on $R^\mathrm{in}$ and $R^\mathrm{up}$, since $R^\mathrm{out}$ and $R^\mathrm{down}$ are obtained from them through the spin-flip conjugation relations~\eqref{eq:Routin} and \eqref{eq:Rdownup}, respectively.

Let us denote by $\eta=|M(\omega-\omega_\mathrm{pole})|$ the dimensionless distance from the pole under consideration.
Here $\omega_\mathrm{pole}$ will be specified for each case below; the same symbol $\eta$ is used throughout this section as a local distance parameter.
By measuring the power-law scaling with respect to $\eta$, we check whether the numerically observed pole order is consistent with the analytic prediction.
In the figures below, the pole is approached along the positive real direction in the complex-$\omega$ plane.
We have also checked the approach from four directions, $\omega=\omega_\mathrm{pole}\pm \xi$ and $\omega=\omega_\mathrm{pole}\pm i \xi$ with $\xi\to 0^+$, and confirmed that the same scaling behavior is obtained for all cases studied in Figs.~\ref{fig:RinRup_MMand0}--\ref{fig:GplusGminusG_Near0_rObs30}.

The numerical result for $R^\mathrm{in}$ shown in the top panel of Fig.~\ref{fig:RinRup_MMand0} exhibits a slope consistent with $-1$, indicating a simple pole at the fundamental Matsubara pole $\omega={}_{-2}\omega^{(-)}_{20}$.
This is consistent with the analytic results obtained above.
Although Fig.~\ref{fig:RinRup_MMand0} shows only the representative case $k=0$, we have checked the same behavior for the Matsubara poles with $0\le k\le 10$.

We have also numerically confirmed the residue formulae~\eqref{eq:Rin-residue} and~\eqref{eq:Rin-residue-MST} in two independent ways: 
by evaluating $(\omega-\omega_\mathrm{pole}) R^\mathrm{in}$ at frequencies sufficiently close to the pole, and by computing the contour integral around the corresponding Matsubara pole in the complex-frequency plane.

Similarly, we numerically analyze $R^{\mathrm{up}}$ in the vicinity of $\omega=0$.
The numerical result for $(s,l)=(-2,2)$ shown in the bottom panel of Fig.~\ref{fig:RinRup_MMand0} exhibits a slope consistent with $-4$, confirming the presence of a fourth-order pole at $\omega=0$, as predicted analytically in \eqref{eq:Ruppole}.
We have also numerically computed $\omega^{l-s} R^{\mathrm{up}}$ for various parameter choices and confirmed that it approaches the leading coefficient ${}_sQ_{lm}(r)$ in \eqref{eq:Rup-leading-coeff-2} as $\omega\to0$.

\subsection{Connection coefficients}
\label{ssec:conn}

We next turn to the connection coefficients, which constitute the second building block entering the Kerr Green's function. 
The connection coefficients $B^\mathrm{inc}$ and $B^\mathrm{ref}$ can be computed by a series expansion in the MST formalism.
Alternatively, we can make use of the connection formula of Heun functions recently obtained~\cite{Bonelli:2022ten}.

As is well known, the zeros of $B^\mathrm{inc}$ correspond to the QNM frequencies and can be computed, for example, by the continued-fraction method.
On the other hand, as discussed in Appendix~\ref{apps:MMcc}, within a fixed asymptotic sector the explicit connection formulae for both $B^\mathrm{inc}$ and $B^\mathrm{ref}$ contain the common factor $\Gamma(\delta)$.
Hence, they develop poles at
$\delta = -k$ with $k=0,1,2,\ldots$,
which precisely coincide with the Matsubara poles ${}_s\omega^{(-)}_{mk}$ given in \eqref{eq:MMfreq}.
The same pole condition can also be derived independently in the MST formalism (see Appendix~\ref{apps:MMcc-mst}).

This is not accidental: $R^\mathrm{in}$ is the local solution around the outer horizon, while $B^\mathrm{inc}$ and $B^\mathrm{ref}$ are the connection coefficients that express this same local solution in the asymptotic basis at infinity.
Therefore, within a fixed asymptotic sector, singularities in the normalization of the local horizon solution are naturally inherited by the corresponding connection coefficients unless canceled by an additional zero in the connection problem.
In this sense, the same pole condition is encoded both in the local horizon solution $R^\mathrm{in}$ and in its connection to the basis at infinity.

As demonstrated in Appendices~\ref{apps:MMcc} and \ref{apps:MMcc-mst}, we can derive semianalytic formulae for the residues of $B^\mathrm{inc}$ and $B^\mathrm{ref}$ at the Matsubara poles, 
\begin{align}
\label{eq:Binc-residue}
\res_{\omega={}_s\omega^{(-)}_{mk}}{}_sB^{\mathrm{inc}}_{lm}
&=
\left.
i\kappa_+\frac{(-1)^k}{k!}
\frac{{}_sB^{\mathrm{inc}}_{lm}}{\Gamma(\delta)}
\right|_{\omega={}_s\omega^{(-)}_{mk}}, \\
\label{eq:Bref-residue}
\res_{\omega={}_s\omega^{(-)}_{mk}}{}_sB^{\mathrm{ref}}_{lm}
&=
\left.
i\kappa_+\frac{(-1)^k}{k!}
\frac{{}_sB^{\mathrm{ref}}_{lm}}{\Gamma(\delta)}
\right|_{\omega={}_s\omega^{(-)}_{mk}} .
\end{align}

On the other hand, for nonzero spin weight $s\ne 0$, the MST low-frequency analysis directly shows that the connection coefficients possess zero-frequency singularities,
\begin{align}
\label{eq:Binczero} {}_sB^{\mathrm{inc}}_{lm} &= \order{\omega^{s-l-1}},
\qquad (\omega\to0), \\
\label{eq:Brefzero} {}_sB^{\mathrm{ref}}_{lm} &= \order{\omega^{-s-l-1}},
\qquad (\omega\to0),
\end{align}
for generic parameters.
For the gravitational lowest multipole $(s,l)=(-2,2)$, this gives a fifth-order pole in $B^\mathrm{inc}$ and a first-order pole in $B^\mathrm{ref}$.

\begin{figure}[t]
\centering
\includegraphics[width=0.95\linewidth]{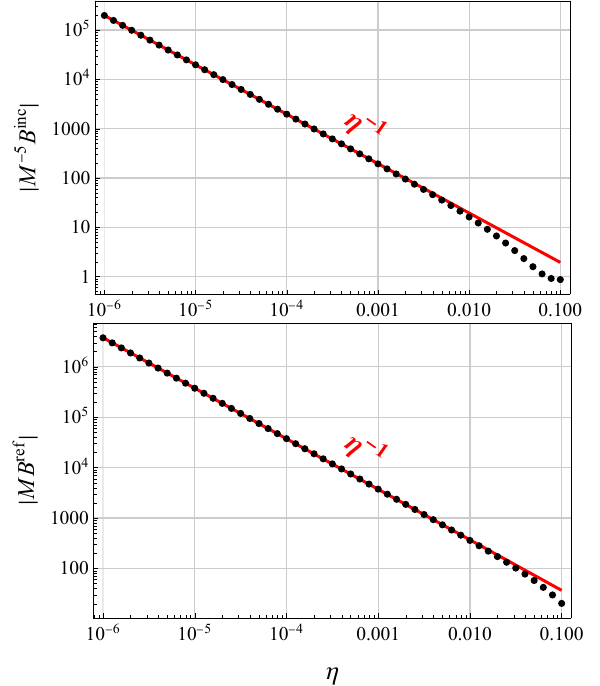}
\caption{
Magnitudes of ${}_sB^{\mathrm{inc}}_{lm}$ (top) and ${}_sB^{\mathrm{ref}}_{lm}$ (bottom) in the vicinity of the fundamental Matsubara pole $\omega={}_s\omega^{(-)}_{m0}$, shown as functions of $\eta=|M(\omega-{}_s\omega^{(-)}_{m0})|$.
Here, $(s,l,m)=(-2,2,2)$ and $a/M=0.7$.
Black dots represent numerical results, while the red lines indicate reference scalings proportional to $\eta^{-1}$.
}
\label{fig:BincBref_MMk=0}
\end{figure}

We now numerically verify these poles and residues for the connection coefficients.
Using the same distance parameter $\eta=|M(\omega-\omega_\mathrm{pole})|$ as in the previous subsection, we present $B^\mathrm{inc}$ and $B^\mathrm{ref}$ near the fundamental Matsubara pole $\omega={}_{-2}\omega^{(-)}_{20}$ in Fig.~\ref{fig:BincBref_MMk=0}, where both quantities exhibit slopes consistent with $-1$, confirming that they possess simple poles at the Matsubara frequency.
As in the case of $R^\mathrm{in}$ and $R^\mathrm{up}$, we have checked the same behavior for the Matsubara poles with $0\le k\le 10$.
We have also numerically confirmed the residue formulae~\eqref{eq:Binc-residue} and \eqref{eq:Bref-residue} by evaluating $(\omega-\omega_\mathrm{pole}) B^\mathrm{inc/ref}$ and by the contour integral around the pole.

\begin{figure}[t]
\centering
\includegraphics[width=0.95\linewidth]{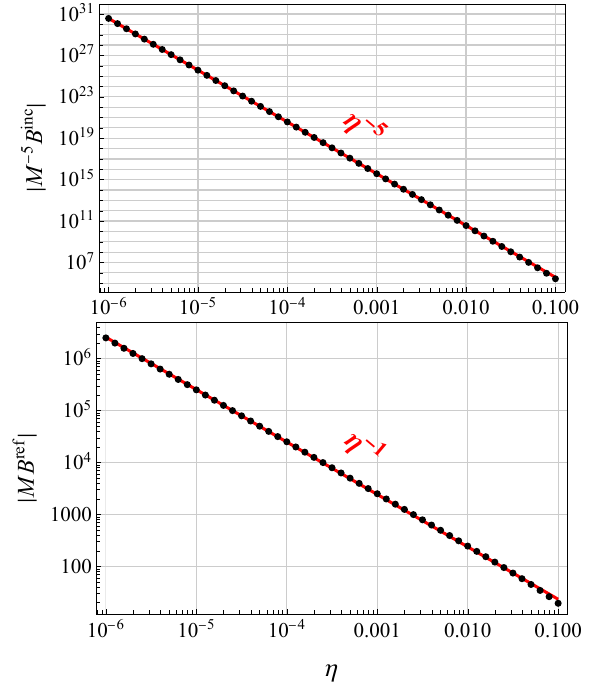}
\caption{
Magnitudes of ${}_sB^{\mathrm{inc}}_{lm}$ (top) and ${}_sB^{\mathrm{ref}}_{lm}$ (bottom) in the vicinity of $\omega=0$, shown as functions of $\eta=|M\omega|$.
Here, $(s,l,m)=(-2,2,2)$ and $a/M=0.7$.
Black dots represent numerical results, while the red lines indicate reference scalings proportional to $\eta^{-5}$ and $\eta^{-1}$, respectively.
}
\label{fig:BincBref_Near0}
\end{figure}

We also numerically confirm the low-frequency behavior of $B^\mathrm{inc}$ and $B^\mathrm{ref}$.
In Fig.~\ref{fig:BincBref_Near0}, we present $B^\mathrm{inc}$ and $B^\mathrm{ref}$ as functions of $\eta=|M\omega|$, which are consistent with \eqref{eq:Binczero} and \eqref{eq:Brefzero}.

\subsection{Green's function}
\label{ssec:Green}

We now combine the previous ingredients into the Green's function itself and clarify how its pole-zero structures involve QNMs, Matsubara modes, and zero-frequency singularities.

The contribution $G^{(+)}$ in \eqref{eq:Gpm} contains the ratio $B^\mathrm{ref}/B^\mathrm{inc}$.
At the level of the local connection formula in a fixed asymptotic sector, both $B^\mathrm{inc}$ and $B^\mathrm{ref}$ contain the same explicit factor $\Gamma(\delta)$.
Therefore, the explicit Matsubara-pole factors associated with $\delta=-k$ cancel in the ratio $B^\mathrm{ref}/B^\mathrm{inc}$ and do not appear explicitly in $G^{(+)}$.
The QNM poles, however, remain: since $G^{(+)}\propto 1/B^\mathrm{inc}$, the zeros of $B^\mathrm{inc}$ generate poles of $G^{(+)}$.

Let us next consider the low-frequency behavior for nonzero spin weight $s\ne 0$.
Using \eqref{eq:Ruppole}, \eqref{eq:Rdownpole}, \eqref{eq:Binczero}, and \eqref{eq:Brefzero}, we obtain from \eqref{eq:Gpm}
\begin{align}
\label{eq:Gpmzero}
G^{(\pm)}
&=\order{\omega^{-2l-1}}.
\end{align}
Thus, both $G^{(+)}$ and $G^{(-)}$ possess the same-order zero-frequency singularity.
For the gravitational lowest multipole $(s,l)=(-2,2)$, this corresponds to a fifth-order zero-frequency pole in each decomposed contribution.

The total Green's function $G$, however, is regular at zero frequency in the present fixed-radius, transmission-normalized radial problem.
Indeed, from the original expression \eqref{eq:Green}, together with the low-frequency dependence of $R^\mathrm{in}$, $R^\mathrm{up}$, and $B^\mathrm{inc}$ clarified above, we obtain 
\be \label{eq:Gzero} G=\order{\omega^0}. \ee
Therefore, the negative-power terms in $G^{(+)}+G^{(-)}$ must cancel collectively, including possible subleading singular terms, leaving the total Green's function regular at $\omega=0$.
The zero-frequency poles shown for $G^{(+)}$ and $G^{(-)}$ in Fig.~\ref{fig:Graph_ComplexPlane} should thus be understood as poles of the decomposed contributions, not as poles of the total Green's function.

\begin{figure}[t]
\centering
\includegraphics[width=0.95\linewidth]{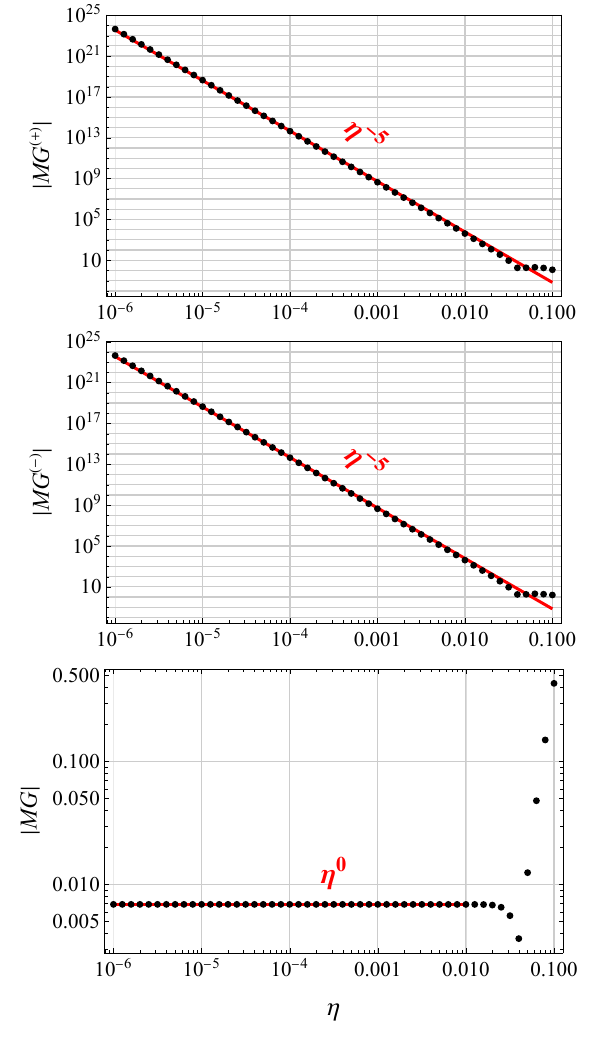}
\caption{
Magnitudes of ${}_sG^{(+)}_{lm}(r,r')$ (top), ${}_sG^{(-)}_{lm}(r,r')$ (middle), and ${}_sG_{lm}(r,r')$ (bottom) in the vicinity of $\omega=0$, shown as functions of $\eta=|M\omega|$.
Here, $(s,l,m)=(-2,2,2)$, $a/M=0.7$, $r/M=30$, and $r'/M=10$.
Black dots represent numerical results, while the red lines indicate reference scalings proportional to $\eta^{-5}$, $\eta^{-5}$ and $\eta^0$, respectively.
}
\label{fig:GplusGminusG_Near0_rObs30}
\end{figure}

We numerically verify these zero-frequency behaviors of the Green's function.
As shown in Fig.~\ref{fig:GplusGminusG_Near0_rObs30}, the numerical results are consistent with the analytical results \eqref{eq:Gpmzero} and \eqref{eq:Gzero}.

The cancellation of the zero-frequency pole in the total Green's function does not necessarily mean that the zero-frequency singularities of the decomposed contributions are irrelevant to the time-domain signal.
As mentioned above, in the early-time regime $|r_*-r'_*|<t-t'<|r_*+r'_*|$, the decomposed components $G^{(+)}$ and $G^{(-)}$ enter the contour analysis in different ways.
Therefore, the zero-frequency poles of $G^{(\pm)}$ may still be important for an analytical understanding of the prompt response.

Note that the above order counting is performed at fixed finite radii.
If the observation point is instead taken to future null infinity $\mathscr I^+$ with the outgoing amplitude normalized there, the small-frequency factor associated with the fixed-radius outgoing solution on the observer side is removed.
Using $R^\mathrm{up}(r)\to r^{-1-2s}e^{i\omega r_*}$ at $\mathscr I^+$ in \eqref{eq:Gpm} while keeping the source point at finite radius, one obtains
\begin{align}\label{eq:Gpmzero_infinity}
G^{(\pm)}\big|_{\mathscr I^+}=\order{\omega^{-s-l-1}}, \qquad
G\big|_{\mathscr I^+}=\order{\omega^0} .
\end{align}
Thus, the zero-frequency order relevant to an asymptotic observer can differ from the fixed-radius sectoral order in \eqref{eq:Gpmzero}.

The same order-counting structure is also found in the RW formulation.
As shown in Appendix~\ref{app:RWMST}, the RW MST representation gives
\begin{align}
G_{\rm RW}^{(\pm)}=\order{\omega^{-2l-1}},
\qquad
G_{\rm RW}=\order{\omega^0}.
\end{align}
Thus, the sectoral zero-frequency singularities of the decomposed Green's function contributions and their cancellation in the total Green's function are not artifacts of the particular Teukolsky master variable or its normalization.

A final remark is that the statements made here concern the pole structure visible in a fixed asymptotic sector of the frequency-domain representation and in the up/down decomposition used in \eqref{eq:Rindecomp} and \eqref{eq:Gpm}.
A different decomposition of the Green's function may redistribute explicit pole factors among the separate contributions, even though the total Green's function is unchanged.
A complete treatment of the inverse-frequency integral, including branch cuts, source terms, possible sector changes of the asymptotic basis at infinity, and alternative decompositions, requires a separate global analysis of the Stokes continuation and is left for future work.

\section{Conclusions and discussion}
\label{sec:con}

We have investigated the pole structure of Kerr black-hole perturbations in the frequency domain, focusing on the building blocks of the Green's function for the radial Teukolsky equation: the homogeneous solutions, the connection coefficients, and the Green's function itself.
We have addressed, from first principles, which singular structures are present in asymptotically flat rotating black holes within the Teukolsky formalism, and how they differ from the structures that have recently been identified in nonrotating or asymptotically de Sitter settings, often in the Regge--Wheeler--Zerilli (RWZ) formalism.
More specifically, we have clarified the Matsubara poles in the homogeneous solutions and connection coefficients, the cancellation of the explicit Matsubara-pole factors in the decomposed Green's function contribution, and the zero-frequency singularities of the homogeneous solutions, connection coefficients, and decomposed Green's function contributions.

Using both the confluent Heun and MST representations of the homogeneous solutions of the radial Teukolsky equation, we showed that the homogeneous solutions and the connection coefficients develop simple poles at the Matsubara frequencies ${}_s\omega^{(\pm)}_{mk}$ identified in \eqref{eq:MMfreq} and~\eqref{eq:MMfreqPlus}, and also verified this behavior numerically.
This establishes the Matsubara pole structure explicitly within the Teukolsky formalism for asymptotically flat subextremal Kerr perturbations, without relying on Kerr/CFT arguments or on a near-extremal approximation.
We also found that, at the level of the local fixed-sector connection formula, the explicit Matsubara-pole factors cancel in the ratio $B^\mathrm{ref}/B^\mathrm{inc}$ entering $G^{(+)}$.
Thus, the Matsubara poles are present in the homogeneous solutions and connection problem, but they are not inherited straightforwardly as poles of the decomposed Green's function contribution.
In the RW formalism for nonrotating black holes, the corresponding MST analysis gives an analogous Matsubara structure, although its detailed form differs because of the choice of master variable and normalization.
This should be contrasted with asymptotically de Sitter analyses in the RWZ formalism, where Matsubara poles can directly organize the early-time response.

Further, using the low-frequency expansion based on the MST formalism, we clarified the zero-frequency scaling of the transmission-normalized homogeneous solutions and connection coefficients.
Consequently, both decomposed Green's function contributions behave as $G^{(\pm)}=\order{\omega^{-2l-1}}$.
For the gravitational lowest multipole $(s,l)=(-2,2)$, each decomposed contribution therefore contains a fifth-order zero-frequency pole.
However, these singularities cancel collectively in the total radial Green's function, which is regular at $\omega=0$ for fixed radii in the present normalization.
This cancellation was also verified numerically.
The RW MST analysis leads to the same order counting for the decomposed Green's function contributions and for the total Green's function, namely, $G_\mathrm{RW}^{(\pm)}=\order{\omega^{-2l-1}}$ and $G_\mathrm{RW}=\order{\omega^0}$.
This agreement suggests that the sectoral zero-frequency singularities and their cancellation in the total Green's function are not artifacts of the particular Teukolsky master variable or its normalization.

Taken together, our results provide a frequency-domain foundation for the study of prompt response in Kerr spacetime.
They show that asymptotically flat rotating black holes possess a richer fixed-sector analytic structure than is visible from the QNM spectrum alone, and identify the homogeneous solutions, connection coefficients, and decomposed Green's function contributions as the basic analytic ingredients for future time-domain analyses.
In particular, the zero-frequency singularities are not poles of the total radial Green's function, but they do appear in the decomposed contributions $G^{(+)}$ and $G^{(-)}$ that enter contour decompositions of the early-time response.
The corresponding sectoral pole order can also depend on whether the Green's function is evaluated at fixed finite radii or at future null infinity with the outgoing amplitude normalized there [compare \eqref{eq:Gpmzero} with \eqref{eq:Gpmzero_infinity}].
It will therefore be important to clarify how these sectoral zero-frequency singularities combine with branch cuts, source terms, observer limits, and contour prescriptions in the actual time-domain waveform.

The higher-order zero-frequency singularities of the decomposed Green's function contributions are also interesting in the broader context of non-Hermitian spectral structures in black-hole perturbation theory.
A pole of order $N$ at $\omega=\omega_0$ would, at the level of an isolated pole contribution to the inverse-frequency integral, generate a contribution of the form $P_{N-1}(t)e^{-i\omega_0 t}$, where $P_{N-1}(t)$ is a polynomial of degree at most $N-1$.
Such polynomial prefactors have recently been highlighted as a non-Hermitian aspect of black-hole perturbations, in connection with the resonant behavior of high-spin Kerr QNMs associated with nearby second-order exceptional-point structures~\cite{Motohashi:2024fwt,Yang:2025dbn,PanossoMacedo:2025xnf}.
In the present problem, the fifth-order zero-frequency poles for the gravitational lowest multipole occur in the decomposed contributions rather than in the total Green's function.
Whether and how these sectoral singularities leave an observable polynomial imprint in prompt response must therefore be analyzed.

The Green's function analysis given here should be extended from the local fixed-sector connection problem and the up/down decomposition to the full global analytic structure, including the Stokes continuation at the irregular singular point and the associated branch-cut structure.
In addition, although the RW comparison in the present work already shows that the Green's function-level low-frequency order counting is not tied to the Teukolsky master variable alone, a more systematic comparison with the RWZ or Sasaki--Nakamura formulations remains to be explored.
In particular, source normalizations, transformation formulae between master variables, and global analytic continuations should be incorporated to determine the representation-independent content of the full time-domain response.
We leave this broader analysis for future work.

\acknowledgments
We thank Emanuele Berti, Kei-ichiro Kubota, Yusuke Manita, and David Pereñiguez for useful discussions.
H.M.\ was supported in part by JSPS KAKENHI Grant No.~JP22K03639 and No.~JP26K21813.

\appendix
\section{Confluent Heun functions}
\label{app:HeunCbasics}

In this appendix, we summarize key properties of the confluent Heun functions~\cite{Heun1889} employed in the main text.
For general background, we refer the reader to the classical textbooks~\cite{ronveaux1995heun,slavianov2000special}, more recent mathematical developments~\cite{Maier_2006,Bonelli:2022ten,Lisovyy:2022flm}, and reviews on physical applications~\cite{Hortacsu:2011rr,Vieira:2016ubt}; for applications to black-hole perturbations in particular, see~\cite{Fiziev:2009wn,CarneirodaCunha:2019tia,Bonelli:2021uvf,Minucci:2024qrn}.

\subsection{Series expansions}

The confluent Heun equation is given by 
\be \frac{d^2w}{dz^2} + \left( \epsilon + \frac{\gamma}{z} + \frac{\delta}{z-1} \right) \frac{dw}{dz} + \frac{\alpha z - q}{z(z-1)} w = 0, \ee
which possesses regular singular points at $z=0,1$ and an irregular singular point at $z=\infty$. 

$\HeunC(q,\alpha,\gamma,\delta,\epsilon;z)$ is the standard confluent Heun function whose Frobenius expansion around $z=0$ is given by
\be \label{eq:HeunC_def} \HeunC(q,\alpha,\gamma,\delta,\epsilon;z)=\sum_{k=0}^\infty c_k z^k, \ee
where the coefficients $c_k$ satisfy the three-term recurrence relation
\begin{equation}\label{eq:HeunC_rec}\begin{split}
& c_{-1}=0, \quad c_{0}=1, \\
& f_k c_{k+1} + g_k c_k + h_k c_{k-1} = 0, 
\end{split}\end{equation}
with
\begin{equation}\label{eq:HeunC_reccoeff}\begin{split}
f_k & = -(k + 1) (k + \gamma), \\
g_k & = k (k - \epsilon + \gamma + \delta - 1) - q, \\
h_k & = (k - 1) \epsilon + \alpha.
\end{split}\end{equation}

On the other hand, $\HeunC_\infty (q,\alpha,\gamma,\delta,\epsilon; z^{-1})$ denotes an asymptotic solution around the irregular singular point $z=\infty$,
\be \label{eq:HeunCinf_def} \HeunC_\infty(q, \alpha, \gamma, \delta, \epsilon; z^{-1}) = \sum_{k=0}^\infty \f{d_k}{z^k}, \ee
where the coefficients $d_k$ satisfy
\begin{equation}\begin{split}
& d_{-1}=0, \quad d_{0}=1, \\
& \tilde f_k d_{k + 1} + \tilde g_k d_k + \tilde h_k d_{k - 1} = 0 ,
\end{split}\end{equation}
with
\begin{equation}\begin{split}
\tilde f_k & = (k + 1) \epsilon, \\
\tilde g_k & = -(k + 1 - \gamma - \delta + \epsilon + \alpha/\epsilon) (\alpha/\epsilon  + k) +q , \\
\tilde h_k & = (\alpha/\epsilon + k - 1)(\alpha/\epsilon + k - \gamma).
\end{split}\end{equation}
Here we have divided the recurrence relation by $\epsilon^2$, so this form assumes $\epsilon\ne0$.
The degenerate limit $\epsilon\to0$
must therefore be treated separately.

\begin{widetext}
\subsection{Independent solutions}

Around each singular point, we have two linearly independent local solutions.
Near $z=0$, they are given by
\begin{align}
\psi_1^{(0)} &= \HeunC(q,\alpha,\gamma,\delta,\epsilon;z), \\
\psi_2^{(0)} &= z^{1-\gamma} \HeunC (q+ (1-\gamma)(\epsilon-\delta),\alpha+ (1-\gamma)\epsilon,2-\gamma,\delta,\epsilon; z) .
\end{align}
Near $z=1$, the two local solutions are
\begin{align}
\label{eq:psi1(1)}
\psi_1^{(1)} &= \HeunC(q-\alpha,-\alpha,\delta,\gamma,-\epsilon; 1-z) , \\
\label{eq:psi2(1)}
\psi_2^{(1)} &= (1 - z)^{1 - \delta}
  \HeunC(q - \alpha + (\gamma + \epsilon) (\delta - 1), -\alpha + (\delta - 1) \epsilon, 2 - \delta, \gamma, -\epsilon; 1 - z) .
\end{align}
Around the irregular singular point $z=\infty$, the two asymptotic solutions are
\begin{align}
\psi_1^{(\infty)} &= z^{-\alpha/\epsilon} \HeunC_\infty (q,\alpha,\gamma,\delta,\epsilon; z^{-1}) , \\
\psi_2^{(\infty)} &= e^{-\epsilon z} z^{\alpha/\epsilon-\gamma-\delta} \HeunC_\infty (q -\gamma\epsilon, \alpha - \epsilon (\gamma+\delta) ,\gamma,\delta,-\epsilon; z^{-1}) .
\end{align}

The irregular singularity at $z=\infty$ is of rank one, and the two asymptotic solutions are distinguished by the exponential factor $e^{-\epsilon z}$.
In the radial Teukolsky problem, \eqref{eq:zr} and \eqref{eq:Heunparams} imply
\begin{align}
\epsilon z = 2\nu_3 z = 2 i \omega (r-r_-) \sim 2 i \omega r,
\qquad (r\to\infty).
\end{align}
The Stokes rays are the directions along which the relative exponential factor changes from decay to growth, and are therefore determined by
\begin{align}
\Re(\epsilon z)=0.
\end{align}
Along the physical radial ray $r>r_+$, this condition reduces to $\Im\omega=0$.
Thus, the upper and lower half-planes of the complex-$\omega$ plane correspond to different asymptotic sectors at infinity, and the identification of the $\HeunC_\infty$ basis with outgoing and ingoing solutions must be understood sectorwise.

Restoring the prefactor in \eqref{eq:Rpsi}, we find, up to algebraic factors,
\begin{align}
R_1^{(\infty)} &\sim e^{+\nu_3 z} \sim e^{+i\omega r}
\qquad (r\to\infty), \\
R_2^{(\infty)} &\sim e^{+\nu_3 z-\epsilon z}=e^{-\nu_3 z}
\sim e^{-i\omega r}
\qquad (r\to\infty).
\end{align}
For real $\omega$, these are naturally identified with outgoing and ingoing behaviors, respectively.
For complex $\omega$, the outgoing and ingoing labels are understood by analytic continuation from real $\omega$.
However, their dominant/subdominant character changes across the Stokes rays: in the upper half-plane $e^{+i\omega r}$ is exponentially suppressed while $e^{-i\omega r}$ is exponentially enhanced, and the opposite holds in the lower half-plane.
Thus, the asymptotic basis at infinity, and hence any connection coefficient involving $\psi^{(\infty)}_{1,2}$, should be understood within a fixed asymptotic sector and for a fixed branch choice.

\subsection{Connection coefficients}

Analytic formulae for connection coefficients of Heun functions and their confluences have recently been obtained in Refs.~\cite{Bonelli:2022ten,Lisovyy:2022flm}.
Here we focus on the connection between the local solution around $z=1$ and the asymptotic basis at $z=\infty$,
\be \psi_1^{(1)} = C_{11}^{(1\infty)} \psi_1^{(\infty)} + C_{12}^{(1\infty)} \psi_2^{(\infty)} . \ee
Through \eqref{eq:InOutUpDown}, this relation corresponds to decomposing the ``in'' solution into the ``up'' and ``down'' solutions.

Reference~\cite{Bonelli:2022ten} gives two asymptotic representations of the connection coefficients, corresponding to the large-$L$ and small-$L$ expansions [Eqs.~(4.2.19) and (4.2.21), respectively].
For the present purpose, we do not need the full expressions.
The essential point is that, in either representation, both connection coefficients contain the common factor $\Gamma(\delta)$.
This shared factor is sufficient to identify the Matsubara pole locations in a fixed asymptotic sector.

There are two qualifications to this statement.
First, an explicit asymptotic connection formula requires a definite choice of representation, asymptotic sector, and branch, as also reflected in the phase ambiguity discussed in Ref.~\cite{Bonelli:2022ten}.
Second, the pole structure visible in such a fixed-sector connection formula is not, by itself, a global statement about the full Green's function.
A global analysis would require tracking the Stokes continuation between the upper and lower half-plane sectors.

Thus, in the confluent Heun analysis below, we use the common factor $\Gamma(\delta)$ only to identify the fixed-sector Matsubara pole structure of the local connection coefficients.
The same pole locations are independently reproduced from the MST expressions for the asymptotic amplitudes in Appendix~\ref{app:mst}, and their realization as poles is also checked numerically in the main text.

\section{Pole structures and residues from the confluent Heun representation}
\label{app:poleres}

In this appendix, we derive the pole and residue formulae used in the main text from the confluent Heun representation.
We first discuss the simple poles of the local confluent Heun function and apply the result to the Matsubara poles of $R^\mathrm{in}$ and $R^\mathrm{out}$.
We then summarize how the common $\Gamma(\delta)$ factor in the Heun connection coefficients gives the corresponding Matsubara poles and residues of the connection coefficients in a fixed asymptotic sector.

\subsection[Matsubara poles of Rin and Rout]{Matsubara poles of $R^\mathrm{in}$ and $R^\mathrm{out}$}
\label{apps:MMRinRout}

First, we derive the pole and residue formula for the local confluent Heun function
$\HeunC(q,\alpha,\gamma,\delta,\epsilon;z)$, which is defined by the Frobenius expansion \eqref{eq:HeunC_def}, with coefficients $c_k$ satisfying the recurrence relation \eqref{eq:HeunC_rec} and \eqref{eq:HeunC_reccoeff}.
For generic $\gamma$, the coefficients $c_k$ are determined recursively.
However, when $\gamma=-n$ with $n=0,1,2,\ldots$, the coefficient $f_n=-(n+1)(n+\gamma)$ vanishes.
As a result, $c_{n+1}$ generically develops a simple pole.
Since, for $k\le n$, the factor $f_{k-1}$ remains nonzero, the coefficients $c_0,c_1,\ldots,c_n$ stay finite as $\gamma\to -n$.
Moreover, the recurrence relation shows inductively that all subsequent coefficients $c_k$ with $k\ge n+1$ inherit the same simple pole.
Equivalently, the possible poles of $c_k$ are contained in the factor $(\gamma)_k^{-1}=[\gamma(\gamma+1)\cdots(\gamma+k-1)]^{-1}$, where the Pochhammer symbol is defined as $(x)_n=\Gamma(x+n)/\Gamma(x)$, and the pole at $\gamma=-n$ appears for $k\ge n+1$.
Therefore, for generic parameters, $\HeunC(q,\alpha,\gamma,\delta,\epsilon;z)$ has simple poles at $\gamma=-n$ with $n=0,1,2,\ldots$.

We now evaluate the residue of the confluent Heun function at $\gamma=-n$.
Since the coefficients $c_k$ with $k\le n$ remain finite, we have
\be
\lim_{\gamma\to -n}(\gamma+n)\HeunC(q,\alpha,\gamma,\delta,\epsilon;z)
= \lim_{\gamma\to -n}(\gamma+n)\sum_{k=n+1}^{\infty} c_k z^k .
\ee
Let us first extract the coefficient of the leading singular term $z^{n+1}$.
Using the recurrence relation at $k=n$, we find
\begin{align}
A_n(q,\alpha,\delta,\epsilon)
\coloneqq
\lim_{\gamma\to -n}(\gamma+n)c_{n+1}
&=
\lim_{\gamma\to -n}(\gamma+n)\frac{g_n c_n+h_n c_{n-1}}{-f_n}
\nonumber\\
&=
\left.
\frac{\bigl[n(-\epsilon+\delta-1)-q\bigr]c_n+\bigl[(n-1)\epsilon+\alpha\bigr]c_{n-1}}{n+1}
\right|_{\gamma=-n} .
\label{eq:An-def}
\end{align}
This is finite because $c_{n-1}$ and $c_n$ are finite at $\gamma=-n$.

Next, define rescaled coefficients as follows:
\be
c'_k \coloneqq \frac{1}{A_n}\lim_{\gamma\to -n}(\gamma+n)c_{k+n+1},
\qquad (k=0,1,2,\ldots) .
\label{eq:ckprime-def}
\ee
By construction,
\be c'_{-1}=0,\qquad c'_0=1 . \ee
From the original recurrence relation~\eqref{eq:HeunC_rec}, we have
\be f_{n+k+1}c_{n+k+2}+g_{n+k+1}c_{n+k+1}+h_{n+k+1}c_{n+k}=0 . \ee
Multiplying this recurrence by $(\gamma+n)/A_n$ and then taking the limit $\gamma\to -n$, we obtain
\be F_k c'_{k+1}+G_k c'_k+H_k c'_{k-1}=0 , \ee
where
\begin{align}\begin{split}
F_k &\coloneqq \left.f_{n+k+1}\right|_{\gamma=-n}
= -(k+1)(k+n+2), \\
G_k &\coloneqq \left.g_{n+k+1}\right|_{\gamma=-n}
= (n+k+1)(k-\epsilon+\delta)-q , \\
H_k &\coloneqq \left.h_{n+k+1}\right|_{\gamma=-n}
= (n+k)\epsilon+\alpha .
\end{split}\end{align}

These coefficients can be rewritten as
\begin{align}\begin{split}
F_k &= -(k+1)(k+\gamma'), \\
G_k &= k(k-\epsilon+\gamma'+\delta-1)-q', \\
H_k &= (k-1)\epsilon+\alpha' ,
\end{split}\end{align}
with the shifted parameters
\be
\gamma' = n+2,\qquad
\alpha'=\alpha+(n+1)\epsilon,\qquad
q'=q+(n+1)(\epsilon-\delta).
\label{eq:shifted-params}
\ee
Therefore, the generating function of $c_k'$ is again a confluent Heun function:
\be
\sum_{k=0}^{\infty} c_k' z^k
=
\HeunC\bigl(q',\alpha',\gamma',\delta,\epsilon;z\bigr) .
\ee

Putting everything together, we finally arrive at
\begin{align}
\lim_{\gamma\to -n}(\gamma+n)\HeunC(q,\alpha,\gamma,\delta,\epsilon;z)
&=
\lim_{\gamma\to -n}(\gamma+n)\sum_{k=n+1}^{\infty} c_k z^k
\nonumber\\
&=
z^{n+1}\sum_{k=0}^{\infty}
\left[\lim_{\gamma\to -n}(\gamma+n)c_{k+n+1}\right] z^k
\nonumber\\
&=
A_n z^{n+1}\HeunC\bigl(q',\alpha',\gamma',\delta,\epsilon;z\bigr) .
\end{align}
Using \eqref{eq:shifted-params}, this becomes
\be \label{eq:HeunC-residue}
\lim_{\gamma \to -n} (\gamma + n) \HeunC(q, \alpha, \gamma, \delta, \epsilon; z) 
=A_n z^{n+1} \HeunC(q + (n + 1) (\epsilon - \delta), \alpha + (n + 1) \epsilon , n + 2, \delta, \epsilon; z), 
\ee
with $A_n=A_n(q,\alpha,\delta,\epsilon)$ given in \eqref{eq:An-def}.
This shows that $\HeunC(q,\alpha,\gamma,\delta,\epsilon;z)$ has simple poles at
$\gamma=-n$, and that the residue is again expressed in terms of a confluent
Heun function with shifted parameters.

Next, to evaluate the residues of $R^\mathrm{in}$ and $R^\mathrm{out}$ at the Matsubara frequencies, we apply the same argument to the local solutions around $z=1$.
The two local solutions $\psi_1^{(1)}$ and $\psi_2^{(1)}$ are given in \eqref{eq:psi1(1)} and \eqref{eq:psi2(1)}, respectively.
For $\psi_1^{(1)}$, the relevant Heun parameter is $\delta$, and the pole is located at $\delta=-k$.
Using \eqref{eq:HeunC-residue}, we obtain
\begin{align}
\res_{\delta=-k}\psi_1^{(1)}
&= A_k(q-\alpha,-\alpha,\gamma,-\epsilon)
(1-z)^{k+1}
\HeunC\bigl(q-\alpha-(k+1)(\gamma+\epsilon),
-\alpha-(k+1)\epsilon,
k+2,
\gamma,
-\epsilon;1-z\bigr).
\label{eq:psi1-residue} 
\end{align}
For $\psi_2^{(1)}$, the relevant Heun parameter is $2-\delta$, and the pole is located at $2-\delta=-k$.
The prefactor $(1-z)^{1-\delta}$ cancels the factor $(1-z)^{k+1}$ generated by \eqref{eq:HeunC-residue}.
We therefore find
\begin{align}
\res_{2-\delta=-k}\psi_2^{(1)}
&= A_k\bigl(q-\alpha+(\gamma+\epsilon)(k+1),
-\alpha+(k+1)\epsilon,
\gamma,
-\epsilon\bigr)
\HeunC\bigl(q-\alpha,
-\alpha,
k+2,
\gamma,
-\epsilon;1-z\bigr) .
\label{eq:psi2-residue}
\end{align}

From \eqref{eq:Rpsi}, \eqref{eq:InOutUpDown}, and 
\be \label{eq:ddelta}
\frac{d\omega}{d\delta}=i\kappa_+,
\ee
we obtain the residues of $R^\mathrm{in}$ and $R^\mathrm{out}$ at the Matsubara frequencies as \eqref{eq:Rin-residue} and \eqref{eq:Rout-residue} in the main text, respectively.

\subsection{Matsubara poles of connection coefficients}
\label{apps:MMcc}

Next, let us consider poles and residues of the connection coefficients in a fixed asymptotic sector.
As discussed in Appendix~\ref{app:HeunCbasics}, the fixed-sector connection coefficients $C_{11}^{(1\infty)}$ and $C_{12}^{(1\infty)}$ contain the common explicit factor $\Gamma(\delta)$.
They therefore exhibit fixed-sector Matsubara poles at
\be
\delta=-k,
\qquad (k=0,1,2,\ldots) .
\ee
Within a fixed sector, the corresponding residues follow immediately from the residue of the Gamma function,
\be \label{eq:Gamma-residue}
\res_{c=-k}\Gamma(c)=\frac{(-1)^k}{k!}.
\ee
Thus,
\begin{align}
\res_{\delta=-k}C_{11}^{(1\infty)}
&=
\left.
\frac{(-1)^k}{k!}
\frac{C_{11}^{(1\infty)}}{\Gamma(\delta)}
\right|_{\delta=-k}, \\
\res_{\delta=-k}C_{12}^{(1\infty)}
&=
\left.
\frac{(-1)^k}{k!}
\frac{C_{12}^{(1\infty)}}{\Gamma(\delta)}
\right|_{\delta=-k} .
\end{align}

The asymptotic amplitudes $B^{\mathrm{inc}}$ and $B^{\mathrm{ref}}$ are obtained from these fixed-sector connection coefficients, up to the normalization factors specified in \eqref{eq:InOutUpDown} and the asymptotic normalization of $R^\mathrm{up}$ and $R^\mathrm{down}$.
Since these additional factors are regular at the Matsubara frequencies, the same $\Gamma(\delta)$ factor gives the residues of the asymptotic amplitudes.
Using \eqref{eq:ddelta}, we find \eqref{eq:Binc-residue} and \eqref{eq:Bref-residue} in the main text.

\section{Pole structures and residues from the MST formalism}
\label{app:mst}

While the local Matsubara pole structure is most transparently exposed in the confluent Heun representation, the same pole locations can also be identified directly in the MST formalism.
This provides an independent consistency check and is particularly useful for analyzing the low-frequency limit.
In this appendix, we summarize the MST results relevant to the Matsubara and zero-frequency pole structures.
For a thorough description of the MST method, we refer the reader to Refs.~\cite{Mano:1996mf,Mano:1996gn,Mano:1996vt,Sasaki:2003xr}.

Throughout this appendix, we adopt unit-transmission normalization.
Also, mostly following Ref.~\cite{Sasaki:2003xr}, we use the notation
\be \label{eq:MSTnotation}
\ve=2M\omega,\quad
\chi=\frac{a}{M},\quad
\kappa=\sqrt{1-\chi^2},\quad
\tau=\frac{\ve-m\chi}{\kappa},\quad
\ve_+=\frac{\ve+\tau}{2}, \quad 
\hat z=\omega(r-r_-), \quad 
\hat z-\ve\kappa=\omega(r-r_+).
\ee
Here, $\ve$ should not be confused with the Heun parameter $\epsilon$.
The symbol $\kappa$ in the MST formalism denotes $\sqrt{1-\chi^2}$ and should not be confused with the horizon surface gravities $\kappa_\pm$ used in the main text.
Although the dimensionless Kerr spin is often denoted by $q$ in the MST literature, we denote it by $\chi$ to avoid confusion with the accessory parameter $q$ of the confluent Heun equation.

For comparison with another master-variable formulation, we discuss the corresponding structures in the RW MST formalism in Appendix~\ref{app:RWMST}.

\subsection[Matsubara poles of Rin and Rout]{Matsubara poles of $R^\mathrm{in}$ and $R^\mathrm{out}$}
\label{apps:MMRinRout-mst}

In the Teukolsky MST formalism, the transmission-normalized ``in'' solution can be written as
\begin{align}
R^\mathrm{in} = \frac{1}{\bar B^\mathrm{trans}} ( K_\nu R_C^\nu + K_{-\nu-1} R_C^{-\nu-1} ),
\label{eq:Rin-MST}
\end{align}
where $\bar B^\mathrm{trans}$ is the transmission amplitude,
\begin{align}
\bar B^\mathrm{trans}
=
\left(\frac{\ve\kappa}{\omega}\right)^{2s}
e^{i\kappa\ve_+(1+\frac{2\ln\kappa}{1+\kappa})}
\sum_{n=-\infty}^{\infty} a_n^\nu ,
\label{eq:Btrans-MST-app}
\end{align}
$R_C^\nu$ is the Coulomb-wave series solution, 
\begin{align}\label{eq:RCnu}
R_C^\nu &= e^{-i\hat z}2^\nu (\ve\kappa)^{-s-i\ve}\hat z^{\nu+i\ve}
\sum_{n=-\infty}^{\infty} \sum_{j=0}^{\infty} D_{n,j} \hat z^{n+j} , \\
D_{n,j} &= \frac{(-1)^n(2i)^{n+j}}{j!} \frac{\Gamma(n+\nu+1-s+i\ve+j)}{\Gamma(2n+2\nu+2+j)} \frac{(\nu+1+s-i\ve)_n}{(\nu+1-s+i\ve)_n} a_n^\nu .
\end{align}
$K_\nu$ is the matching coefficient, 
\begin{align}\label{eq:Knu}
K_\nu
&=
e^{i\ve\kappa}(2\ve\kappa)^{s-\nu-N}2^{-s} i^N
\frac{\Gamma(1-s-2i\ve_+)\Gamma(N+2\nu+2)}
{\Gamma(N+\nu+1-s+i\ve)\Gamma(N+\nu+1+i\tau)\Gamma(N+\nu+1+s+i\ve)}
\notag\\
&\quad\times
\left[
\sum_{n=N}^{\infty}
\frac{(-1)^n\Gamma(n+N+2\nu+1)}{(n-N)!}
\frac{\Gamma(n+\nu+1+s+i\ve)}{\Gamma(n+\nu+1-s-i\ve)}
\frac{\Gamma(n+\nu+1+i\tau)}{\Gamma(n+\nu+1-i\tau)}
a_n^\nu
\right]
\notag\\
&\quad\times
\left[
\sum_{n=-\infty}^{N}
\frac{(-1)^n}{(N-n)!(N+2\nu+2)_n}
\frac{(\nu+1+s-i\ve)_n}{(\nu+1-s+i\ve)_n}
a_n^\nu
\right]^{-1},
\end{align}
$a_n^\nu$ satisfy the three-term recurrence relation
\be \label{eq:anu_rec} \alpha_n^\nu a_{n+1}^\nu + \beta_n^\nu a_n^\nu + \gamma_n^\nu a_{n-1}^\nu=0 , \ee
with
\begin{align}\begin{split}
\alpha_n^\nu &= \frac{i\ve\kappa(n+\nu+1+s+i\ve)(n+\nu+1-i\ve)(n+\nu+1+i\tau)}{(n+\nu+1)(2n+2\nu+3)}, \\
\beta_n^\nu &= -\lambdabar-s(s+1)+(n+\nu)(n+\nu+1)+\ve^2+\ve(\ve-m\chi)+\frac{\ve(\ve-m\chi)(s^2+\ve^2)}{(n+\nu)(n+\nu+1)}, \\
\gamma_n^\nu &= -\frac{i\ve\kappa(n+\nu-s+i\ve)(n+\nu-s-i\ve)(n+\nu-i\tau)}{(n+\nu)(2n+2\nu-1)},
\end{split}\end{align}
and $\nu$ denotes the renormalized angular momentum, which is determined by the condition that the above three-term recurrence relation admits a minimal solution.
Here, the integer $N$ appearing in \eqref{eq:Knu} is the arbitrary MST matching index.
The full matching coefficient is independent of this choice, although individual factors in the expression are not.
In the following analysis, we are therefore free to fix this integer, and we take $N=0$ for convenience.

From the matching coefficients $K_\nu$ and $K_{-\nu-1}$ in \eqref{eq:Knu}, $R^\mathrm{in}$ contains the explicit gamma-function factor $\Gamma(1-s-2i\ve_+)$.
Using the definitions~\eqref{eq:MSTnotation}, one finds $1-s-2i\ve_+=\delta$, given in \eqref{eq:Heunparams}.
Therefore, for generic parameters, it has simple poles when $\delta=-k$ with $k=0,1,2,\ldots$.
This reproduces the Matsubara frequencies ${}_s\omega^{(-)}_{mk}$ identified in \eqref{eq:MMfreq} from the confluent Heun analysis.
The corresponding residue formula~\eqref{eq:Rin-residue-MST} follows from the residue of the gamma function~\eqref{eq:Gamma-residue} and from \eqref{eq:ddelta}. 

The solution $R^\mathrm{out}$ is obtained from $R^\mathrm{in}$ through the spin-flip conjugation relation~\eqref{eq:Routin}.
Hence, we obtain the other Matsubara branch ${}_s\omega^{(+)}_{mk}$ in \eqref{eq:MMfreqPlus}, corresponding to $1+s+2i\ve_+ = 2-\delta = -k$ with $k=0,1,2,\ldots$.
The corresponding residue is then given by \eqref{eq:Rout-residue-MST}. 

\subsection{Matsubara poles of connection coefficients}
\label{apps:MMcc-mst}

The transmission-normalized asymptotic amplitudes are given by
\begin{align}
B^\mathrm{inc}
&=\frac{1}{\bar B^\mathrm{trans}}
\omega^{-1}
\left[
K_\nu
- i e^{-i\pi\nu}
\frac{\sin \pi(\nu-s+i\ve)}{\sin \pi(\nu+s-i\ve)}
K_{-\nu-1}
\right]
A_+^\nu e^{-i(\ve\ln\ve - \frac{1-\kappa}{2}\ve)},
\label{eq:Binc-MST-app}
\\
B^\mathrm{ref}
&=\frac{1}{\bar B^\mathrm{trans}}
\omega^{-1-2s}
\left[
K_\nu + i e^{i\pi\nu} K_{-\nu-1}
\right]
A_-^\nu e^{i(\ve\ln\ve - \frac{1-\kappa}{2}\ve)},
\label{eq:Bref-MST-app}
\end{align}
where 
\begin{align}
A_+^\nu
&=
e^{-\frac{\pi}{2}\ve}
e^{\frac{i \pi}{2}(\nu+1-s)}
2^{-1+s-i\ve}
\frac{\Gamma(\nu+1-s+i\ve)}{\Gamma(\nu+1+s-i\ve)}
\sum_{n=-\infty}^{\infty} a_n^\nu,
\label{eq:Aplus-MST-app}
\\
A_-^\nu
&=
2^{-1-s+i\ve}
e^{-\frac{i \pi}{2}(\nu+1+s)}
e^{-\frac{\pi}{2}\ve}
\sum_{n=-\infty}^{\infty}
(-1)^n
\frac{(\nu+1+s-i\ve)_n}{(\nu+1-s+i\ve)_n}
a_n^\nu .
\label{eq:Aminus-MST-app}
\end{align}

Both $K_\nu$ and $K_{-\nu-1}$ contain the same explicit gamma-function factor $\Gamma(1-s-2i\ve_+)=\Gamma(\delta)$.
For generic parameters, this factor produces poles at the Matsubara frequencies.
The remaining factors in \eqref{eq:Binc-MST-app} and \eqref{eq:Bref-MST-app} are regular at these frequencies.
Thus, $B^\mathrm{inc}$ and $B^\mathrm{ref}$ have simple poles at the Matsubara frequencies ${}_s\omega^{(-)}_{mk}$ given in \eqref{eq:MMfreq}.
Because the pole-producing factor is identical to the one found in the confluent Heun analysis, the corresponding residues reduce to \eqref{eq:Binc-residue} and \eqref{eq:Bref-residue}, respectively.

The above analysis shows that the Matsubara pole structures and residues identified in the confluent Heun formulation are reproduced independently in the MST formalism.

\subsection{Low-frequency scaling}
\label{apps:zerofreqpole}

We now turn to the low-frequency power counting based on the MST expansion~\cite{Mano:1996vt}.
In this subsection, we restrict the zero-frequency order counting to nonzero spin weight $s\ne 0$, which includes the gravitational case $s=-2$ studied in the main text. 
The scalar case $s=0$ is exceptional because leading terms in the MST sums can cancel, and is not included in the order estimates below.

We first note that the renormalized angular momentum $\nu$ satisfies
\be \label{eq:nu_w=0} \nu=l+\order{\ve^2}. \ee
For fixed $n$ away from the resonant negative-index sector, the recurrence relation~\eqref{eq:anu_rec} gives the low-frequency scaling
\be a_n^\nu=\tilde\alpha_n\,\ve^{|n|}+ \order{\ve^{|n|+1}}, \ee
where $\tilde\alpha_n$ obeys 
\begin{align}
\label{eq:alpha_n_recurrence_epsilon}
\begin{split}
\tilde{\alpha}_0 &= 1,
\\
\tilde{\alpha}_n
&=
\frac{i\,(n+l-s)^2\bigl[(n+l)\kappa+i m \chi\bigr]}
{n(n+l)(n+2l+1)(2n+2l-1)}\,\tilde{\alpha}_{n-1},
\qquad (n\ge 1),
\\
\tilde{\alpha}_n
&=
-\frac{i\,(n+l+1+s)^2\bigl[(n+l+1)\kappa-i m \chi\bigr]}
{n(n+l+1)(n+2l+1)(2n+2l+3)}\,\tilde{\alpha}_{n+1},
\qquad (n\le -1).
\end{split}
\end{align}
This recurrence gives
\begin{align}
\tilde{\alpha}_n
= \begin{cases} 
\displaystyle \left(\frac{i\kappa}{2}\right)^n
\frac{(l+1-s)_n^2
\left(l+1+\frac{i m\chi}{\kappa}\right)_n}
{n!\,(l+1)_n\,(2l+2)_n\left(l+\frac12\right)_n}, 
& (n\ge 0), \\
\displaystyle \left(\frac{i\kappa}{2}\right)^{|n|}
\frac{(-l-s)_{|n|}^2
\left(-l+\frac{i m\chi}{\kappa}\right)_{|n|}}
{|n|!\,(-l)_{|n|}\,(-2l)_{|n|}\left(-l-\frac12\right)_{|n|}}, 
& (n\le -1).
\end{cases}
\label{eq:alpha_n_closed}
\end{align}
This fixed-index expansion is nonuniform in the resonant sector $n\le -l-1$, where $\nu-l=\order{\ve^2}$ must be retained before taking the low-frequency limit.
Consequently, it cannot by itself be used to assign orders to the individual raw MST normalization factors.

The transmission-normalized solutions $R^\mathrm{in}$ and $R^\mathrm{out}$ are regular at $\omega=0$ for fixed $r>r_+$.
For the normalized connection coefficients, direct MST evaluation gives
\begin{align}
\label{eq:BincBref-zero-scaling}
B^\mathrm{inc} =\order{\omega^{s-l-1}}, \qquad
B^\mathrm{ref} =\order{\omega^{-s-l-1}}.
\end{align}
These orders are also confirmed by the numerical results in Fig.~\ref{fig:BincBref_Near0}.

For completeness, the MST representation of $R^\mathrm{up}$ is
\begin{align}
R^\mathrm{up}
&=
\frac{1}{\bar C^\mathrm{trans}} 2^\nu e^{-\pi\ve} e^{-i\pi(\nu+1+s)}
e^{i\hat z}\hat z^{\nu+i\ve_+}(\hat z-\ve\kappa)^{-s-i\ve_+} \notag\\
&\quad\times\sum_{n=-\infty}^{\infty}
\frac{(\nu+1+s-i\ve)_n}{(\nu+1-s+i\ve)_n}
a_n^\nu (2i\hat z)^n
U\!\left(n+\nu+1+s-i\ve,\,2n+2\nu+2,\,-2i\hat z\right).
\label{eq:MST-up}
\end{align}
Here, $U(a,b,z)$ is Tricomi's confluent hypergeometric function, which is also denoted as $\Psi(a,b,z)$ in the literature. 
As stressed in the main text, note that we define $R^\mathrm{up}$ with the normalization by the transmission coefficient $\bar C^\mathrm{trans}$, which is defined by
\begin{align}\label{eq:Ctra} 
\bar C^\mathrm{trans}
&= \omega^{-1-2s}e^{i(\ve\ln\ve-\frac{1-\kappa}{2}\ve)}A_-^\nu .
\end{align}

The raw coefficient $A_-^\nu$ in \eqref{eq:Ctra} is subject to the same resonant-sector qualification stated above.
A uniform analytic treatment of these raw MST factors is beyond the scope of the present work.
The normalized solution itself has been independently checked numerically, and its leading low-frequency form is
\begin{align}
\label{eq:Rup-leading}
{}_sR^\mathrm{up}_{lm}(r)
&= {}_sQ_{lm}(r) \omega^{s-l} + \order{\omega^{s-l+1}},
\end{align}
where
\begin{align}
\label{eq:Rup-leading-coeff}
{}_sQ_{lm}(r)
&=
(-1)^s i^{\,l+s} 2^{s-l}
\frac{(l-s)!}{(l+s)!}
\left(\frac{r-r_+}{r-r_-}\right)^{\frac{i m \chi}{2\kappa}}
\frac{(r-r_+)^{-s}}{(r-r_-)^{\,l+1}}
\sum_{n=0}^{\infty}
\frac{(2n+2l)!}{(n+l-s)!}
\left(\frac{-iM}{r-r_-}\right)^n
\tilde{\alpha}_n .
\end{align}
Substituting the closed form of $\tilde\alpha_n$ in \eqref{eq:alpha_n_closed}, 
we obtain the simplified expression
\begin{align}
\label{eq:Rup-leading-coeff-2}
{}_sQ_{lm}(r)
&=
(-1)^s i^{\,l+s}2^{s-l}
\frac{(2l)!}{(l+s)!}
\left(\frac{r-r_+}{r-r_-}\right)^{\frac{i m\chi}{2\kappa}}
\frac{(r-r_+)^{-s}}{(r-r_-)^{l+1}}
{}_2F_1\left(l+1-s,l+1+\frac{i m\chi}{\kappa};2l+2;\frac{r_+-r_-}{r-r_-}\right).
\end{align}

We can obtain a similar expression for $R^\mathrm{down}_{lm}$ through \eqref{eq:Rdownup}:
\begin{align} \label{eq:Rdown-leading}
{}_sR^\mathrm{down}_{lm}(r)&= \Delta^{-s}(r) \bigl[{}_{-s}Q^*_{lm}(r) \omega^{-s-l} + \order{\omega^{-s-l+1}}\bigr].
\end{align}
Using $\left[{}_2F_1(a,b;c;\frac{r_+-r_-}{r-r_-})\right]^*={}_2F_1(a,b^*;c;\frac{r_+-r_-}{r-r_-})$ since $0<\frac{r_+-r_-}{r-r_-}<1$ for $r>r_+$ and the Euler transformation formula for the hypergeometric function, 
${}_2F_1(a,b;c;z)=(1-z)^{c-a-b}{}_2F_1(c-a,c-b;c;z)$,
we can rewrite the leading-order coefficient as
\begin{align}
\label{eq:Rdown-leading-coeff}
\Delta^{-s}{}_{-s}Q_{lm}^*
=
2^{-l-s}(-i)^{l+s}
\frac{(2l)!}{(l-s)!}
\left(\frac{r-r_+}{r-r_-}\right)^{\frac{i m\chi}{2\kappa}}
\frac{(r-r_+)^{-s}}{(r-r_-)^{l+1}}
{}_2F_1\left(
l+1-s,\,
l+1+\frac{i m\chi}{\kappa};\,
2l+2;\,
\frac{r_+-r_-}{r-r_-}
\right).
\end{align}

Let us remark that the above results \eqref{eq:BincBref-zero-scaling}, \eqref{eq:Rup-leading}, and \eqref{eq:Rdown-leading} are compatible with the decomposition \eqref{eq:Rindecomp} and the regularity of $R^\mathrm{in}$, but this compatibility should not be understood as a statement that follows from the leading powers alone.
Using $R^\mathrm{up}=\order{\omega^{s-l}}$ and $R^\mathrm{down}=\order{\omega^{-s-l}}$, the two products $B^\mathrm{ref}R^\mathrm{up}$ and $B^\mathrm{inc}R^\mathrm{down}$ may each contain singular terms starting at $\order{\omega^{-2l-1}}$.
However, the left-hand side of \eqref{eq:Rindecomp} is the unit-transmission-normalized horizon solution, which is independently seen from the MST representation \eqref{eq:Rin-MST} to be regular at $\omega=0$ for fixed $r>r_+$.
Therefore, the negative-power terms in the sum $B^\mathrm{ref}R^\mathrm{up}+B^\mathrm{inc}R^\mathrm{down}$ must cancel collectively, order by order, leaving the regular solution $R^\mathrm{in}=\order{\omega^0}$.
A direct verification of the individual cancellation coefficients would require the corresponding higher-order low-frequency expansion of all factors and is not needed for the pole-counting argument used here.

\section{Regge--Wheeler comparison in the MST formalism}
\label{app:RWMST}

It is useful to compare the Teukolsky results obtained above with the MST representation of the RW equation~\cite{Mano:1996mf}.
This comparison makes clear which features of the Matsubara and zero-frequency structures are tied to the choice of master variable and normalization, and which survive at the level of Green's function order counting.
The comparison in this appendix concerns the gravitational Regge--Wheeler master variable. It is therefore not affected by the exceptional scalar Teukolsky case $s=0$, which is outside the scope of the low-frequency order counting in Appendix~\ref{app:mst}.

In this appendix, we repeat the main steps of the Teukolsky MST analysis of Appendix~\ref{app:mst} in the RW MST formalism: we identify the Matsubara pole locations of the horizon-adapted solutions and connection coefficients, discuss their cancellation in the decomposed Green's function contribution, and then examine the low-frequency scaling of the homogeneous solutions, connection coefficients, and decomposed Green's function contributions.
As we shall see below, the local Matsubara frequencies and the zero-frequency orders of individual homogeneous solutions are representation dependent.
By contrast, the cancellation of the explicit Matsubara-pole factors in the decomposed Green's function contribution, the sectoral Green's function scaling $G^{(\pm)}=\order{\omega^{-2l-1}}$, and the cancellation giving $G=\order{\omega^0}$ are reproduced in the RW formulation.
Below, we denote the RW master variable by $X$, use $\ve=2M\omega$, and introduce $x=1-r/(2M)$ and $z=\omega r$.

\subsection{Matsubara poles in the RW formalism}

The solution satisfying the ingoing boundary condition at the horizon can be written as
\begin{align}
X^\nu_\mathrm{in}
&=e^{i\ve(x-1)}(-x)^{-i\ve}(1-x)^{-1}
\sum_{n=-\infty}^{\infty}a_n^\nu p_{n+\nu}(x),
\label{eq:RW-Xin-MST}
\end{align}
where
\begin{align}
p_{n+\nu}(x)
&=
\frac{\Gamma(n+\nu-1-i\ve)\Gamma(-n-\nu-2-i\ve)}{\Gamma(1-2i\ve)}
{}_2F_1(n+\nu-1-i\ve,-n-\nu-2-i\ve;1-2i\ve;x).
\label{eq:RW-p-MST}
\end{align}
The horizon-adapted solution $X_\mathrm{in}^\nu$ is not normalized to unit transmission at the horizon.  
Its horizon amplitude is
\be \label{eq:BtransRW}
\mathcal B_\mathrm{trans}^{\rm RW}
= e^{-i\ve}
\sum_{n=-\infty}^{\infty}a_n^\nu
\frac{\Gamma(n+\nu-1-i\ve)\Gamma(-n-\nu-2-i\ve)}
{\Gamma(1-2i\ve)} .
\ee
Thus, if one adopts the same unit-transmission normalization as in the Teukolsky analysis in the main text, one should consider
\be \hat X_\mathrm{in}^\nu
= \frac{X_\mathrm{in}^\nu}{\mathcal B_\mathrm{trans}^{\rm RW}}. \ee
After this normalization, the common overall factor $1/\Gamma(1-2i\ve)$ cancels between $X_\mathrm{in}^\nu$ and $\mathcal B_\mathrm{trans}^{\rm RW}$.

The normalized solution $\hat X_\mathrm{in}^\nu$ is then expressed as a series of hypergeometric functions ${}_2F_1(a,b;c;x)$ with the common parameter $c=1-2i\ve$.
Therefore, unless canceled in the full MST series, it is expected to develop simple poles at $1-2i\ve=-k$, with $k$ being a non-negative integer.
Equivalently, since $\kappa_+=1/(4M)$ for Schwarzschild, the corresponding pole frequencies are
\be \label{eq:RW-MM-candidate} \omega=-i(k+1)\kappa_+, \qquad (k=0,1,2,\ldots). \ee

Similarly, one can consider the solution satisfying the outgoing boundary condition at the horizon and normalize it by its transmission coefficient to obtain $\hat X_\mathrm{out}^\nu$.
The corresponding normalized solution is composed of hypergeometric functions ${}_2F_1(a,b;c;x)$ with the common parameter $c=1+2i\ve$, leading to
\be \label{eq:RW-MM-candidate-plus} \omega=+i(k+1)\kappa_+, \qquad (k=0,1,2,\ldots). \ee
This is consistent with the complex-conjugation structure between $\hat X_\mathrm{in}^\nu$ and $\hat X_\mathrm{out}^\nu$.

We should nevertheless keep in mind that the RW MST solutions also admit equivalent representations in other bases, such as the $X_0^\nu$ and $X_0^{-\nu-1}$ basis~\cite{Mano:1996mf}.
Thus, the pole locations inferred above should be understood as those visible in the present unit-transmission-normalized RW MST representation.
A complete pole-zero analysis of the full RW or RWZ Green's function would further require tracking possible cancellations among homogeneous solutions, connection coefficients, and Wronskian factors.

The same normalization also determines the Matsubara structure of the RW connection coefficients at infinity.
In the notation of Ref.~\cite{Mano:1996mf}, the unnormalized horizon-adapted solution has the asymptotic form
\begin{align}
X^\nu_{\rm in} = A^\nu_{\rm out} e^{iz}z^{i\ve} + A^\nu_{\rm in} e^{-iz}z^{-i\ve}.
\end{align}
These amplitudes themselves do not contain an explicit $\Gamma(1-2i\ve)$ pole factor.
However, the connection coefficients for the unit-transmission-normalized solution are
\begin{align}
\hat A^\nu_{\rm out} = \frac{A^\nu_{\rm out}}{\mathcal B_\mathrm{trans}^{\rm RW}},
\qquad
\hat A^\nu_{\rm in} = \frac{A^\nu_{\rm in}}{\mathcal B_\mathrm{trans}^{\rm RW}}.
\end{align}
Since $\mathcal B_\mathrm{trans}^{\rm RW}$ contains the common factor
$1/\Gamma(1-2i\ve)$, it has zeros at $1-2i\ve=-k$ for generic parameters.
Accordingly, the normalized coefficients $\hat A^\nu_{\rm out}$ and $\hat A^\nu_{\rm in}$ have simple poles at these frequencies.
These are precisely the RW Matsubara frequencies~\eqref{eq:RW-MM-candidate}.

The same pole-producing factor appears in both normalized connection coefficients.
Therefore, in the RW Green's function decomposition, the explicit Matsubara-pole factors cancel in the ratio of connection coefficients $\hat A_{\rm out}/\hat A_{\rm in}$, in direct analogy with the Teukolsky result discussed in \S\ref{ssec:Green}.
Thus, as in the Teukolsky case, the Matsubara poles are visible in the homogeneous horizon-adapted solutions and in their connection coefficients, but are not inherited straightforwardly as explicit poles of the decomposed Green's function contribution.

Let us compare the RW Matsubara frequencies~\eqref{eq:RW-MM-candidate} and \eqref{eq:RW-MM-candidate-plus} with the Teukolsky Matsubara frequencies~\eqref{eq:MMfreq} and \eqref{eq:MMfreqPlus}.
In the Schwarzschild limit of the Teukolsky result, the real shift $m\Omega_+$ vanishes, consistently with the purely imaginary RW frequencies and with the absence of $m$ dependence in the RW equation.
The remaining difference is the spin-weight dependence.
Indeed, the RW locations coincide with the Teukolsky formula only after formally setting $s=0$.
This should not be interpreted as saying that the RW variable describes a scalar field; rather, it reflects that the local pole locations of individual homogeneous solutions depend on the choice of master variable and normalization.
For gravitational perturbations, the Teukolsky variables with $s=\pm2$ and the RW variable are related by nontrivial transformations, such as Chandrasekhar--Sasaki--Nakamura or Darboux transformations, which can shift the apparent local pole structure of the homogeneous solutions.
A full derivation of how the $s$-dependent Teukolsky Matsubara factors are reorganized under such transformations is beyond the scope of the present work.
Here we only note that the RW locations are independent of the Teukolsky spin weight because the RW master variable is used, while the Green's function-level low-frequency order counting below agrees with the Teukolsky result.

\subsection{Low-frequency scaling in the RW formalism}

The zero-frequency behavior of the RW upgoing solution can also be estimated in the MST formalism.
We use the Coulomb-wave representation of the solution satisfying the outgoing condition at infinity~\cite{Mano:1996mf},
\be X_{\rm up}^\nu = \frac{\Gamma(\nu-1-i\ve)\Gamma(\nu+1-i\ve)}{\Gamma(\nu+1+i\ve)\Gamma(\nu+3+i\ve)} X_{C\mathrm{out}}^\nu ,\ee
where 
\begin{align}
X_{C\mathrm{out}}^\nu
&=e^{iz}z^{\nu+1}\left(1-\frac{\ve}{z}\right)^{-i\ve}
2^\nu e^{-\pi\ve}e^{-i\pi(\nu+1)}
\notag\\
&\quad\times
\sum_{n=-\infty}^{\infty}
\frac{\Gamma(n+\nu-1-i\ve)\Gamma(n+\nu+1-i\ve)}
{\Gamma(n+\nu+1+i\ve)\Gamma(n+\nu+3+i\ve)}
a_n^\nu (-2iz)^n
U(n+\nu+1-i\ve,2n+2\nu+2,-2iz) .
\label{eq:RW-XCout-MST}
\end{align}
The prefactor relating $X_{\rm up}^\nu$ to $X_{C\mathrm{out}}^\nu$ is regular and nonzero at $\ve=0$ for integer $l\ge2$, since $\nu=l+\order{\ve^2}$.
Therefore, it does not affect the low-frequency scaling.
This remains true if one normalizes the RW ``up'' solution to unit outgoing amplitude at infinity, since the required normalization is regular at $\ve=0$.

Let us take the limit $\omega\to0$ at fixed $r$.
We can identify the leading-order contribution by the same type of calculation as in Appendix~\ref{apps:zerofreqpole}.
Using $\nu=l+\order{\ve^2}$ and $a_n^\nu=\order{\ve^{|n|}}$, the singular part of the Tricomi function behaves as
\begin{align}
U(n+\nu+1-i\ve,2n+2\nu+2,-2iz)
\simeq z^{-2n-2\nu-1},
\end{align}
whenever this singular term is present with a finite coefficient.
Therefore, each term for which this singular Tricomi contribution is present scales as
\begin{align}
a_n^\nu z^{\nu+1+n} U
&= \order{\omega^{|n|-n-l}}
=
\begin{cases}
\order{\omega^{-l}}, & (n\ge 0),\\
\order{\omega^{-l+2|n|}}, & (n<0).
\end{cases}
\end{align}
If the singular Tricomi term is absent for some negative $n$, the remaining regular term is less singular and does not change the leading order.
Thus, barring an accidental cancellation of the leading sum over $n\ge0$, the unit-outgoing RW up solution behaves as $\omega^{-l}$.
By the conjugation relation, the same scaling applies to the unit-ingoing RW ``down'' solution.
Taken together, the leading low-frequency behavior at fixed radius is
\begin{align}
\hat X_\mathrm{up}^\nu(r)
=\order{\omega^{-l}},
\qquad 
\hat X_\mathrm{down}^\nu(r)
=\order{\omega^{-l}}.
\label{eq:RW-Xup-zero}
\end{align}

This is the RW counterpart of the zero-frequency singularity of the Teukolsky ``up'' and ``down'' solutions, but the pole order differs from the Teukolsky result ${}_sR^\mathrm{up}_{lm}=\order{\omega^{s-l}}$ and ${}_sR^\mathrm{down}_{lm}=\order{\omega^{-s-l}}$ because the master variable and normalization are different.
For the ``up'' solutions, the RW variable is normalized by a standard outgoing amplitude at infinity, $X_\mathrm{up}\simeq e^{i\omega r_*}$, whereas the Teukolsky variable is normalized by the spin-weight-dependent asymptotic factor $R^\mathrm{up}\simeq r^{-1-2s}e^{i\omega r_*}$.
Consequently, there is no direct RW analog of the Teukolsky factor $\bar C^\mathrm{trans}=\order{\omega^{-1-2s}}$ that shifts the zero-frequency pole order.
The low-frequency singularity of $X_\mathrm{up}$ instead comes from the small-argument behavior of the outgoing Coulomb-wave, or equivalently spherical-wave, basis at fixed radius.
Thus, the zero-frequency pole order of a homogeneous radial solution should not by itself be regarded as representation independent; a representation-independent comparison should be made at the level of the full Green's function, including the source normalization and the transformation between the Teukolsky, RWZ, and Sasaki--Nakamura variables.

The RW Green's function decomposition can be analyzed by the same low-frequency order-counting method used for the Teukolsky case in Appendix~\ref{apps:zerofreqpole}.
The horizon amplitude $\mathcal B_\mathrm{trans}^{\rm RW}$ in \eqref{eq:BtransRW} is regular and nonzero at $\omega=0$ for the full MST series, although this regularity is not manifest term by term because individual gamma functions may be singular.
Thus, the unit-transmission-normalized horizon solution satisfies
\begin{align}
\hat X_{\rm in}=\order{\omega^0},
\qquad
\hat X_{\rm out}=\order{\omega^0}.
\end{align}
On the other hand, the asymptotic amplitudes $A^\nu_{\rm in}$ and $A^\nu_{\rm out}$ given in Ref.~\cite{Mano:1996mf} contain the matching coefficient $K_\nu$, whose leading low-frequency behavior is $K_\nu=\order{\omega^{-l-1}}$ for a fixed matching index.
The terms proportional to $K_{-\nu-1}$ are subdominant.
Therefore, after unit-transmission normalization, the RW connection coefficients scale as
\begin{align}
\hat A_{\rm in}=\order{\omega^{-l-1}},
\qquad
\hat A_{\rm out}=\order{\omega^{-l-1}}.
\end{align}
Together with $\hat X_{\rm up}=\order{\omega^{-l}}$ and $\hat X_{\rm down}=\order{\omega^{-l}}$, this gives
\begin{align}
G_{\rm RW}^{(\pm)}=\order{\omega^{-2l-1}},
\qquad
G_{\rm RW}=\order{\omega^0},
\end{align}
in agreement with the Teukolsky order counting for the Green's function contributions, \eqref{eq:Gpmzero} and \eqref{eq:Gzero}.

The fixed-radius order counting above should be distinguished from the order relevant to an observer at future null infinity.
At $\mathscr I^+$, the outgoing solution is normalized as $\hat X_\mathrm{up}^\nu\to e^{i\omega r_*}$, so the factor $\omega^{-l}$ in \eqref{eq:RW-Xup-zero} from the observer side is absent.
Keeping the source point at finite radius, one therefore obtains
\begin{align}
G_{\rm RW}^{(\pm)}\big|_{\mathscr I^+}=\order{\omega^{-l-1}},
\qquad
G_{\rm RW}\big|_{\mathscr I^+}=\order{\omega^0},
\end{align}
in agreement with the post-Minkowskian result at $\mathscr I^+$~\cite{DeAmicis:2026tus}.

\end{widetext}

\bibliography{ref}
\end{document}